# Silicon nitride integrated photonics from visible to mid-infrared spectra

*Kirill A. Buzaverov,[1,2] Aleksandr S. Baburin,[1,2] Evgeny V. Sergeev,[1] Sergey S. Avdeev,[1] Evgeniy S. Lotkov,[1] Sergey V. Bukatin[1], Ilya A. Stepanov[1], Aleksey B. Kramarenko,[1] Ali Sh. Amiraslanov,[1] Danil V. Kushnev,[1] Ilya A. Ryzhikov,[1,3] and Ilya A. Rodionov[1,2,*]*

K.A. Buzaverov, A.S. Baburin, E.V. Sergeev, S.S. Avdeev, E.S. Lotkov, S.V. Bukatin, I.A. Stepanov, A.B. Kramarenko, A.Sh. Amiraslanov, D.V. Kushnev, I.A. Ryzhikov, I.A. Rodionov
FMN Laboratory, Bauman Moscow State Technical University, Moscow 105005, Russia
Email: irodionov@bmstu.ru

K.A. Buzaverov, A.S. Baburin, I.A. Rodionov
Dukhov Research Institute of Automatics (VNIIA), Moscow 127055, Russia

I.A. Ryzhikov
Institute for Theoretical and Applied Electromagnetics Russia Academy of Science, Moscow 125412, Russia

**Abstract:** Recently, silicon nitride ($Si_3N_4$) photonic integrated circuits (PICs) are of a great interest due to their extremely low waveguides losses. The number of $Si_3N_4$ integrated photonics platform applications is constantly growing including the Internet of Things (IoT), artificial intelligence (AI), light detection and ranging (LiDAR) devices, hybrid neuromorphic and quantum computing. Their heterogeneous integration with a III–V platform leads to a new advanced large-scale PICs with thousands of elements. Here, we review key trends in $Si_3N_4$ integrated circuits technology and fill an information gap in the field of state-of-the-art photonic devices operating from visible to mid-infrared spectra. A comprehensive overview of $Si_3N_4$ integrated circuits microfabrication process details (deposition, lithography, etching, etc.) is introduced. Finally, we point out the limits and challenges of silicon nitride photonics performance in a ultrawide range providing routes and prospects for their future scaling and optimization.

## 1. Introduction

Integrated $Si_3N_4$ photonic devices were first demonstrated in the 1990s [1]. In the last three decades, $Si_3N_4$ circuits have been significantly optimized in terms of waveguide losses and scalability. Despite the limitations in the active element fabrication and modulation rates, $Si_3N_4$ remains the most optimal platform for manufacturing passive components for most applications because it has a wide transmission window from 400 to 4000 nm and is characterized by the lowest losses to date at the average level of mode confinement [2]. The disadvantages of $Si_3N_4$ circuits have been successfully overcome by hybrid or heterogeneous integration with photonic circuits based on indium phosphide (InP) [3] and silicon-on-insulator (SOI) [4]. Ecosystems with process design kits (PDKs), multi-project wafer (MPW) runs, packaging, and assembly technologies have already been implemented by stepper-lithography-based fabs, ensuring the rapid development of the PICs market. Several key applications of the $Si_3N_4$ platform in visible (VIS), near-infrared (NIR), short-wavelength infrared (SWIR) and mid-infrared (MWIR) bands are summarized in **Figure 1**.

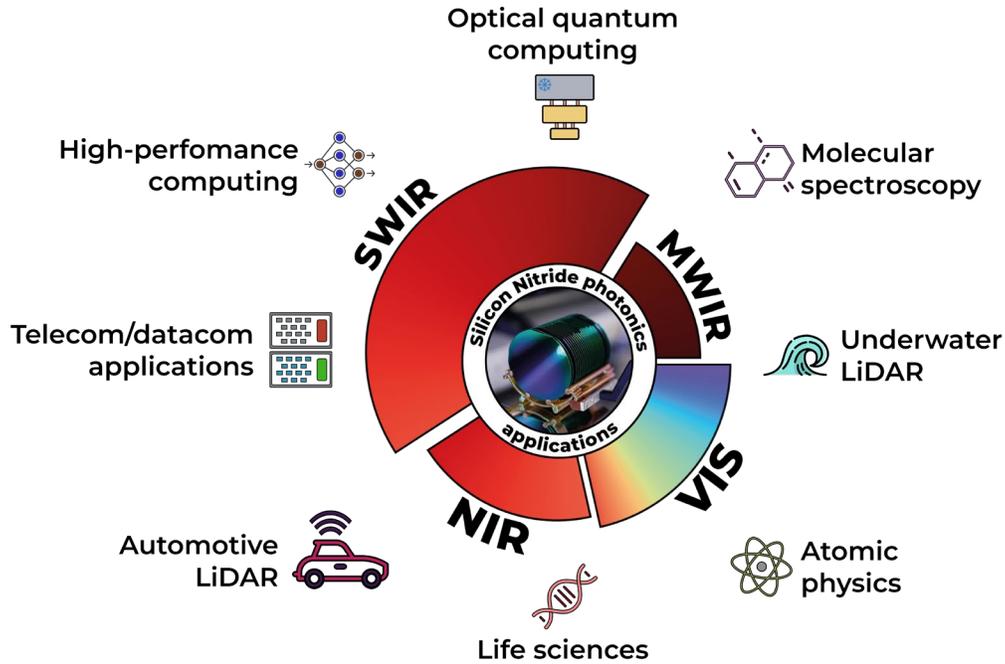

**Figure 1**. Ultra-wideband applications of silicon nitride integrated photonics.

As a key technology, PICs play a critical role in implementing signal processing in the C-band, which is explained by progress in telecom/datacom applications. Currently, C-band circuits are used in quantum processors with 8 [5], 12 [6], and 20 qumodes [7], ultralow-noise highly coherent integrated laser sources [2,8], optical frequency comb generation [9,10], and optical phased array beam scanners for LiDAR systems [11,12]. Previous reviews have mostly focused on the specifics of $Si_3N_4$ devices operating in this wavelength range [13–16]. Recently, Poon et al. reported a review on foundry-compatible platforms mostly for visible spectrum [17]. However, a comprehensive view of $Si_3N_4$ technologies and applications over a wide wavelength range is lacking with regard to the PIC-based devices designed for the next-generation industrial revolution, including, but not limited to, high-sensitivity sensors [18–21], point-of-care diagnostics [22], ultrahigh-resolution spectroscopy [23], quantum processors compatible with state-of-the-art single-photon sources [24], and underwater LiDAR [25].

This article aims to fill the information gap in applications of $Si_3N_4$ photonics, as well as show relevant progress in technology for VIS, NIR, SWIR and MWIR bands. In Section 2, we deeply overview and introduce the $Si_3N_4$ circuits microfabrication process. The details of material deposition, lithography, etching, and coupling structures fabrication are analyzed. While the recent technological progress remains significant, the currently achieved loss in waveguides (WG) and microresonators quality factors operating in a wide wavelength range are presented in Section 3 of the review, along with the latest achievements in creating state-of-the-art devices based on $Si_3N_4$ photonics. Finally, Section 4 summarizes the current state of silicon nitride photonics and discusses challenges for further scaling.

## 2. Silicon nitride photonics materials and technologies

Before discussing the technological advances in the $Si_3N_4$ PICs fabrication, it is worth noting that $Si_3N_4$ waveguides parameters, such as core width and height, dramatically affect device performance in terms of production (**Figure 2**).

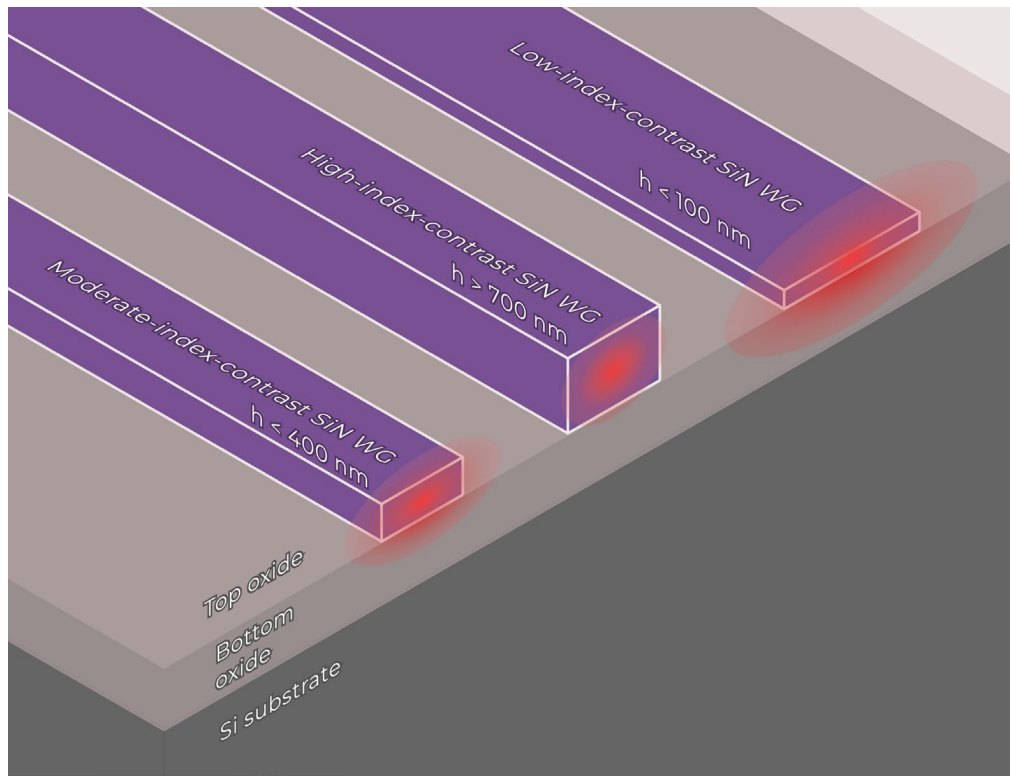

**Figure 2**. Common types of silicon nitride waveguides and typical optical mode localization (photonic integrated circuits scaling potential).

Waveguides with a $Si_3N_4$ core thickness below 400 nm (moderate-index-contrast) are the most universal waveguide type. Relatively high mode localization makes it possible to design sub-micrometer structures with sharp bends [26,27] and provides total absence of losses on higher-order mode excitement [13,14]. However, the propagation loss level in such waveguides is far from desired (typically, 0.1–0.3 dB/cm in the telecommunication bands) due to strong and frequent effective refractive index perturbations in the sidewall roughness [28,29]. Waveguide thicknesses greater than 700 nm (high index contrast) are required to fabricate structures for nonlinear optics applications such as optical frequency comb generation and optical parametric amplification. With the dispersion engineering approach [30–33], the requirements for anomalous group-velocity dispersion are met. The level of 0.01 dB/cm propagation losses is state-of-the-art for such a design because of the lower overlap with the waveguide sidewall roughness. The support of higher-order modes is a disadvantage of thick $Si_3N_4$ waveguides, which makes the routing between straight and bent waveguides challenging [34].

Although propagation losses in the 0.01–0.10 dB/cm range are sufficient for proof-of-concept and small batch devices, certain applications such as quantum technologies [35,36] and neuromorphic computing [37,38] might require even lower waveguide losses. In this case, $Si_3N_4$ waveguides with thicknesses below 100 nm (low index contrast) can be introduced. These high-aspect-ratio waveguides with widths of several micrometers exhibit propagation losses below 0.001 dB/cm owing to low mode localization and almost zero scattering losses caused by minimal overlap with the surface roughness. However, scaling appears to be a common problem in such designs, as the minimum feasible bend radius is on the order of several millimeters or higher.

Despite the importance of waveguide geometry, the main contribution to achieve a minimal level of loss is the technology performance. For almost 30 years of $Si_3N_4$ PICs technology evolution, several manufacturing pathways have been developed, resulting in ultralow propagation losses in high-density photonic networks.

One of the fabrication approaches, the photonic Damascene process, was introduced in 2016 [39]. This process makes it possible to fabricate $Si_3N_4$ circuits with waveguide core thicknesses of up to 1.5 μm due to predefined patterning in the $SiO_2$ preform for stress management [40]. Extremely long and high-temperature reflow of the bottom silicon dioxide and high-precision chemical mechanical planarization were used to obtain the sub-nm level sidewall roughness and state-of-the-art propagation losses of 3 dB/cm (0.4 million Q) at 462 nm [25], 0.01 dB/cm (30 million Q) at 1550 nm [9], and 0.05 dB/cm at 2000 nm [41]. Recently, this technology has been complemented by the implantation of erbium ions, showing a broadband on-chip net gain of 30 dB with propagation loss of 0.05 dB/cm in the C-band [42]. However, the preform reflow stage could lead to silicon diffusion in $SiO_2$ and silicon cluster formation in $SiO_2$ and $Si_3N_4$, resulting in degradation of the optical properties of both materials [43,44]. Thicker layers of silicon dioxide are required to avoid excess absorption loss [45]. In addition, such high-temperature treatment could cause the appearance of transition metals (Cr, Fe, Cu) impurities and enhance their diffusive redistribution [46], resulting in wavelength-independent absorption losses. Considering the above limitations, as well as the high demand for chemical mechanical planarization, the Damascene reflow process is not easily implemented in R&D laboratories for experimental and small batches of devices.

An alternative approach, called the classic subtractive fabrication process, can be considered as the basic production technique for various waveguide core geometries. Although most research groups have used this process in integrated photonics, the Michal Lipson Nanophotonics Group made the main contribution to its design and development in early 2009 [47]. With stepwise optimization of the separate fabrication steps, including electron beam lithography, reactive ion etching, cladding deposition, and chemical mechanical polishing, state-of-the-art propagation losses of 4.6 dB/cm (0.21 million Q) at 530 nm [48], 0.03 dB/cm at 1310 nm [49], and 0.004 dB/cm (67 million Q) at 1550 nm [50] were achieved. The introduction of a specific technique for $Si_3N_4$ low-pressure chemical vapor deposition, including stress-release trenches scribing in buried oxide and multilayer deposition with thermal cycling, makes it possible to easily manufacture waveguide cores with a thickness of up to 1 μm [14].

Even lower losses values are obtained using TriPleX® [51] fabrication approach. A single-stripe geometry with high aspect ratio waveguides ensured a record ultralow propagation loss of 0.09 dB/cm (9.5 million Q) at 461 nm [52], 0.01 dB/cm at 920 nm [53], and 0.00034 dB/cm (720 million Q) at 1570 nm [2], but scalability is dramatically limited. TriPleX waveguides fabrication is a variation of the classic subtractive fabrication process; however, there are several differences. The low-mode confinement factor of the waveguide provides insignificant interaction with the surface roughness but requires a very thick bottom oxide (up to 15 μm) and limits the critical bending radius (up to 1 cm), thus limiting the scalability of the technology [54]. In addition to single-stripe waveguides, symmetric or asymmetric double-stripe TriPleX waveguides (SDS and ADS, respectively) have also been reported [5,7,24,55]. A symmetric double-stripe layout is typically used in components that require tight bending radii and large birefringence. To apply low-loss coupling to the standard single-mode fibers (SMF-28), both $Si_3N_4$ stripes are adiabatically tapered, resulting in fiber-chip coupling losses of <0.5 dB/facet (90% coupling efficiency). The asymmetric double-stripe geometry is highly suitable for combining regions with low and high effective index regions on a single chip, which makes it possible to potentially combine ultralow-loss properties with small bending radii [55].

In summary, all the mentioned above fabrication approaches share several technological operations, and their optimization is quite essential for manufacturing structures with high optical properties, low roughness, and minimal losses. In the next sections, we discuss the key technological operations for $Si_3N_4$ PICs fabrication, focusing on the common processes for R&D and small batch production.

### 2.1. Bottom oxide synthesis

The initial basis for the fabrication of common $Si_3N_4$ PICs is the formation of a high-quality bottom silicon dioxide layer on a Si wafer. This bottom cladding layer determines device operation and prevents high losses owing to substrate leakage with perfect mode isolation at a properly designed thickness.

Silicon oxidizes in a simple form, even at room temperature, when the silicon surface is exposed to oxygen or air. The thin layer of native silicon dioxide has a final thickness of approximately 1–2 nm because the oxygen atoms do not possess enough energy at room temperature to diffuse through the already formed silicon oxide layer. To ensure the high optical quality of the silicon dioxide layer for PICs fabrication, silicon is commonly oxidized in high-temperature furnaces using high-purity grade oxygen, pure pyrogenic or deionized water steam, which are known as dry oxidation and wet oxidation processes, respectively, by the following chemical reactions:

$$\text{Dry oxidation: } Si + O_2 \rightarrow SiO_2 \qquad (1)$$

$$\text{Wet oxidation: } Si + 2H_2O \rightarrow SiO_2 + 2H_2 \qquad (2)$$

The growth time for the required thickness as well as oxide quality depends on the temperature, pressure, and silicon wafer parameters. The Deal–Grove model can be used to calculate the oxidation time for the required thickness [56].

Dry oxidation of silicon provides the densest film with the smoothest possible interface between silicon and the future-deposited core silicon nitride layer. However, hundreds of hours are required for the oxidation of at least 2 μm, which does not fit any time or thermal budget. In contrast, wet oxide growth provides a much faster oxidation rate without significant limitations on film density and interface smoothness. Dry and wet oxidation techniques combination can be used to fabricate thick high-quality oxide layers within a reasonable time (**Figure 3**) [14].

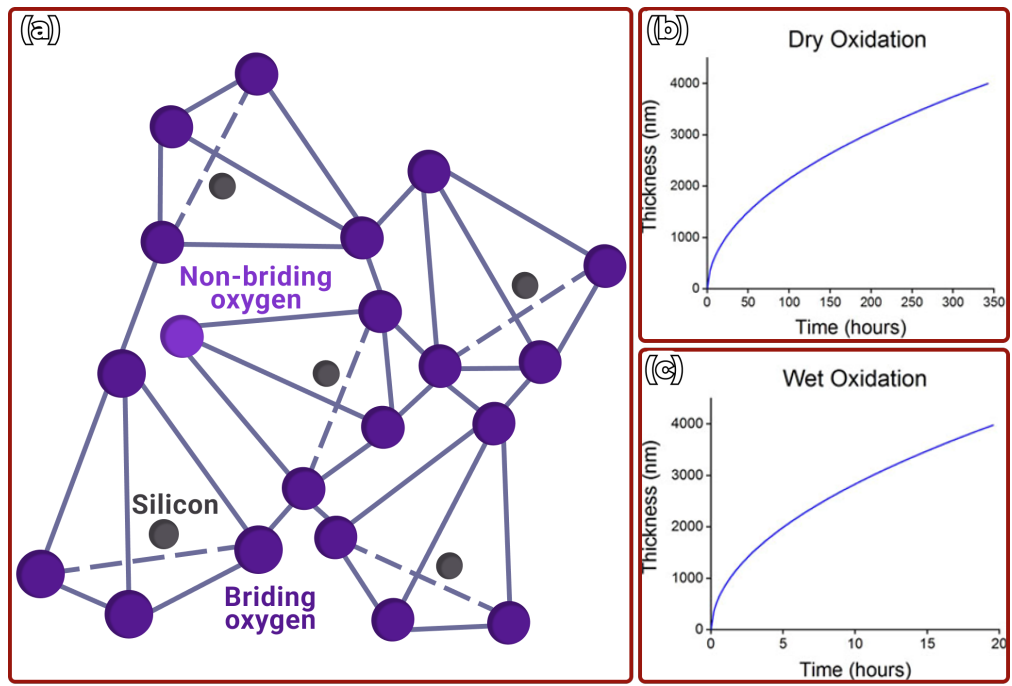

**Figure 3**. Silicon oxidation process: (a) simplified $SiO_2$ molecule structure; silicon dioxide thickness as a function of time for: (b) dry oxidation and (c) wet oxidation, reprinted from [14], under a Creative Commons license (https://creativecommons.org/licenses/by/4.0/).

For certain applications where grating couplers and superconducting nanowire single-photon detectors with high efficiency are required, distributed Bragg reflectors (DBR) can be a proper solution [57,58]. Typical Bragg reflectors films are deposited using low-pressure chemical vapor deposition (LPCVD) and plasma-enhanced CVD (PECVD) methods [59,60]. The chemical vapor deposition of silicon dioxide is a much more complicated process than thermal oxidation. The chemical vapor deposition process stages include the following: (1) transport of the precursors in the gaseous form to the wafer proximity; (2) gas reaction to form a range of daughter molecules; (3) transport of these reactants to the wafer surface; (4) surface reactions; (5) desorption and transport of the by-products from the wafer surface to the exhaust [61]. The classical LPCVD process occurs in hot-wall horizontal or vertical reactors at low pressures (0.1–1.0 Torr) and high temperatures (700–900 °C). Radio-frequency plasma is used in the PECVD process as an alternative energy source, allowing the process to proceed at lower substrate temperatures (250–400 °C) and comply with the standard CMOS manufacturing process. Silane, dichlorosilane (DCS), tetraethoxysilane (TEOS), and nitrous oxide ($N_2O$) are commonly used as precursors in both LPCVD and PECVD. However, silicon dioxide deposited with LPCVD and especially with PECVD techniques exhibits lower density and higher concentration of hydrogen (depending on the chosen precursor) compared to the thermally grown oxide, causing excess absorption and scattering losses [62–64]. In rare cases, highest quality and perfectly conformal SiO$_2$ layer could be deposited with ion-beam deposition. This technique allows for compatibility with the biofunctionalization process and protection of waveguide facets for testing and measurement of photonic devices [65,66].

## 2.2. Silicon nitride deposition

The deposition of the silicon nitride core waveguide layer follows the bottom oxide layer synthesis. The losses and optical performance of future devices are largely influenced by Si$_3$N$_4$ film characteristics; thus, a carefully optimized deposition process is required. Silicon nitride films for PICs are commonly deposited using LPCVD and PECVD. The standard LPCVD Si$_3$N$_4$ process requires high temperatures (around 800°C) and commonly uses ammonia (NH$_3$) and dichlorosilane (DSC) as the nitrogen and silicon sources, respectively, which react according to the following equation:

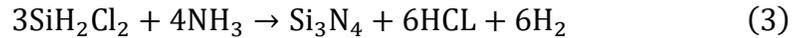

$$3\text{SiH}_2\text{Cl}_2 + 4\text{NH}_3 \rightarrow \text{Si}_3\text{N}_4 + 6\text{HCL} + 6\text{H}_2 \qquad (3)$$

The quality and stress of the deposited silicon nitride obtained using LPCVD depend on the deposition temperature, pressure, gas ratio (DCS/NH$_3$), total gas flow and film thickness. However, when the reaction is stoichiometric, the deposition rate increases (typically to nearly 5 nm/min), preventing more hydrogen from escaping from the silicon nitride film during deposition. The hydrogen escape mechanism relies on thermal activation energy, which enables silicon and nitrogen to be disposed of the silicon nitride film surface via atomic surface migration phenomena, while compelling hydrogen to escape the film. In contrast, if the deposition rate decreases to allow more hydrogen to escape the film, the stress increases. This is a common problem, particularly for thick Si$_3$N$_4$ layers [45,67,68]. To deposit relatively thick Si$_3$N$_4$ films without cracks while minimizing the absorption related to the hydrogen content, an optimum deposition rate should be found. Typically, a DCS/NH$_3$ gas ratio of 0.2–0.3 is used with a deposition rate of nearly 2 nm/min. It minimizes the absorption related to the hydrogen content of the film and allows the deposition of relatively thick silicon nitride films without cracks [69].

As mentioned above, stoichiometric Si$_3$N$_4$ deposition with a film thickness of more than 400 nm is quite challenging owing to high tensile stress, which leads to film cracking or excess wafer bow and production waste. Despite this, waveguides with anomalous group-velocity dispersion are required for the entire range of applications in nonlinear optics [9,10,41,50,70,71], where a stoichiometric Si$_3$N$_4$ core thickness of more than 700 nm is required. Moreover, thicker Si$_3$N$_4$ core is required to fully confine optical mode inside waveguide for mid-IR wavelengths [72]. Several technological methods for stress management have been developed in recent years, including thermal cycling [67], wafer rotation [33,45,69], and filler pattern design [73,74]. In addition to technological solutions, silicon-rich Si$_3$N$_4$ film synthesis could be a useful method for engineering waveguide properties. An increase in the Si content results in a much lower tensile stress and allows the deposition of thicker films [75]. Moreover, alteration in the precursor ratio causes an increase in the refractive index from 2.0 to 2.2 for LPCVD [76] and to 3.1 for PECVD [77]. Owing to an increase in the linear and nonlinear losses in silicon-rich films, bandgap engineering with process conditions tailored for the desired film composition and optical properties is also required (**Figure 4**) [76].

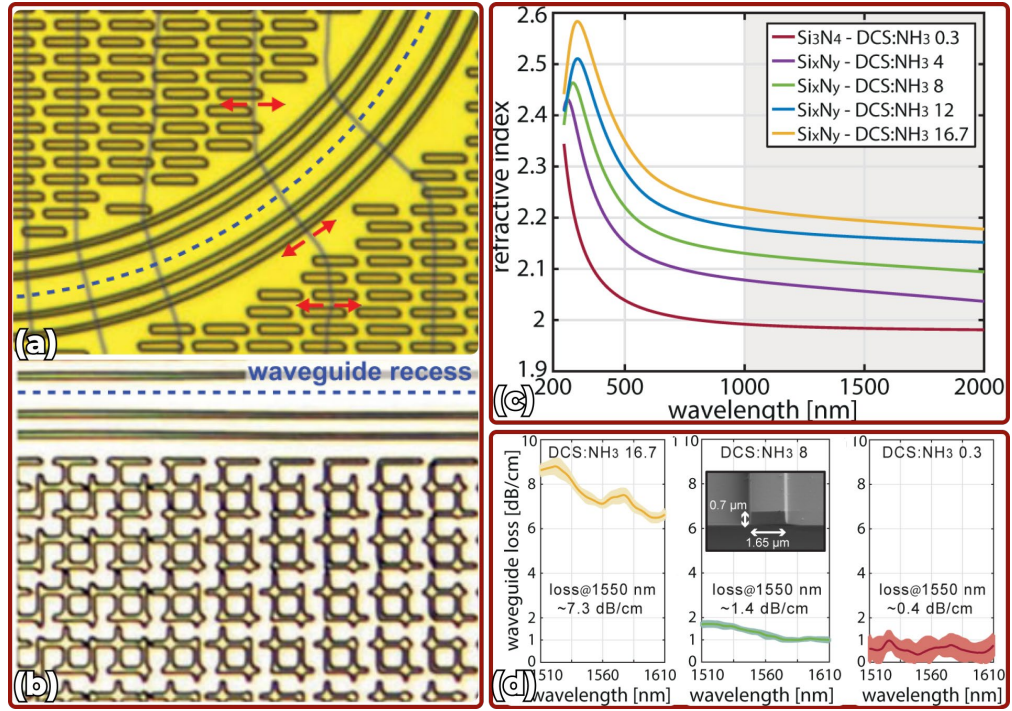

**Figure 4**. Silicon nitride deposition process: (a) crack formation after deposition of thick $Si_3N_4$ film and (b) crack free deposition of thick Si3N4 film with optimized filler pattern, adapted with permission from [73], Copyright (2018) IEEE; (c) silicon nitride refractive index and (d) waveguide propagation loss as a function of different gas composition, adapted with permission from [76], Copyright (2017) Optical society of America.

To ensure compliance with front-end CMOS processing, where an optical circuit needs to be completely manufactured within the back-end process, plasma-enhanced CVD $Si_3N_4$ synthesis appears to be a relevant technology [78–80]. This material is frequently used in microelectronics as an isolation or chemical barrier layer. However, its application in integrated photonics is quite limited because of the higher hydrogen concentration and lower surface quality, which leads to increased absorption losses in the telecommunication band and scattering losses. Further development of complex devices, where microelectronics and PICs are combined, would require additional optimization of PECVD $Si_3N_4$ technology to leverage lower losses over a wide wavelength range [81–86]. Low-temperature reactive sputtering could also be used as an alternative $Si_3N_4$ deposition technique compatible with front-end silicon electronics [87–89]. Magnetron sputtering technology is intrinsically hydrogen free, thereby avoiding optical losses in the telecommunication bands through N–H or Si–H bonds. Moreover, silicon nitride films up to 8 μm thick could be synthesized without crack formation using stress management techniques during sputtering [90,91]. However, this technique has not been widely tested for waveguide fabrication and requires further investigation.

## 2.3. Lithography

The next fabrication step is photonic integrated circuit topology patterning. Ultraviolet (UV) photolithography and electron-beam (e-beam) lithography are the primary methods used in PIC manufacturing. A resolution of a lithography process determines the minimum feature size that can be reliably patterned [92]. It is typically characterized by a parameter known as the minimum feature size. As lithography techniques advance, smaller sizes can be achieved, leading to high-resolution structures. Smaller minimum feature sizes enable the fabrication of narrower and more precise elements, which can enhance the photonic devices performance by mitigating propagation losses and enabling a tighter light confinement.

UV photolithography is an essential technique for photonics and advanced semiconductor microelectronics manufacturing. It offers improved resolution and enhanced pattern transfer capabilities, enabling the fabrication of smaller and more intricate devices. Deep ultraviolet (DUV) photolithography steppers are used in modern production facilities for the PICs fabrication, especially in telecommunication-driven applications [9,93,94]. This technology is related to projection lithography and requires costly photomasks which contains the exact photonic circuit design transferred to photoresist via ultraviolet light exposure through the projection lens system. Fast pattern transfer is one of the key advantages of DUV photolithography, allowing the exposure of the entire wafer within seconds. However, the patterned topology and minimum feature size are strongly limited by photomasks design and technology.

Another common technique is e-beam lithography (EBL), where patterns can be directly written on a substrate without the need for masks. This direct writing capability enables quick prototyping and facilitates the fabrication of customized designs [95]. EBL allows the creation of arbitrary patterns and complex geometries [96]. It offers the freedom to design and fabricate structures that are not achievable with other lithography techniques, thus providing greater flexibility for research and development [97,98]. Extremely high EBL system resolution allows the fabrication of dense waveguide patterns with a sidewall roughness of less than 1 nm and sub-50 nm features, e.g., inversed tapers, directional couplers [99]. However, an e-beam lithography throughput is significantly lower compared to photolithography. It results in many hours exposure time and depends on writing conditions: pattern size, beam current and dose factor. Large-area high-resolution nanostructures patterned with mask-less methods such as electron-beam lithography suffer from stitching errors [100,101]. Its addressable area is limited by a maximum beam deflection area, which defines the writing field (WF) size. Patterns exceeding the dimensions of the writing field must be stitched together by sample mechanical repositioning. The state-of-the-art EBL tools provide sub-10 nm stitching error values [102], which could significantly influence propagation losses of patterned waveguides. For example, a 2 dB/cm loss contribution from stitching was reported for suspended MIR slot waveguides with a stitching error of 28 nm [103]. Several reasons are considered to cause stitching errors in e-beam lithography [100]. In the ideal case, the EBL tool stage shifts along one coordinate axis in the pass direction, and the topology shift is exactly equal to the exposure field size. However, in real cases, there are several EBL tool errors resulting in degraded field stitching accuracy.

The most significant errors are shift errors, scanning field distortions, scanning field rotations, and deflection scale errors. The shift error is stochastic, and its amplitude depends on the stage mechanical quality and parameters of the stage-beam feedback loop, as well as on the stability of the electron beam position in the column between cyclic calibration procedures. The last three errors are more systematic and can be predominantly corrected by the tool. The most promising method for reducing stochastic stitching errors is the multi-pass exposure technique [100]. It is based on overlapping the working fields in a certain proportion. In the EBL, it is implemented by an appropriate exposure dose in the overlapped areas using a series of partial-dose exposures. This method can significantly reduce the mean error over the chip [99], but it is time-consuming. The next technique to decrease stochastic errors, which could be induced by an improperly aligned writing field, is stitching error compensation [104] (Figure 5). Errors originating from the imperfect calibration of WF are systematic; for example, stitching errors are identical at the boundaries of all WFs. In this case, specific values of the errors can be pre-measured on a test sample; then, the adjacent WFs are precisely shifted during device manufacture according to the premeasured stitching error. Another advanced technique to eliminate the influence of systematic and stochastic errors involves adiabatically tapering at the field boundaries to reduce the propagation loss caused by misalignment [105] (**Figure 5**). Thus, a preliminary design modification should be performed to use adiabatic structures. To use them, the waveguide is tapered wider at the lithographic field boundaries to increase robustness against stitching errors.

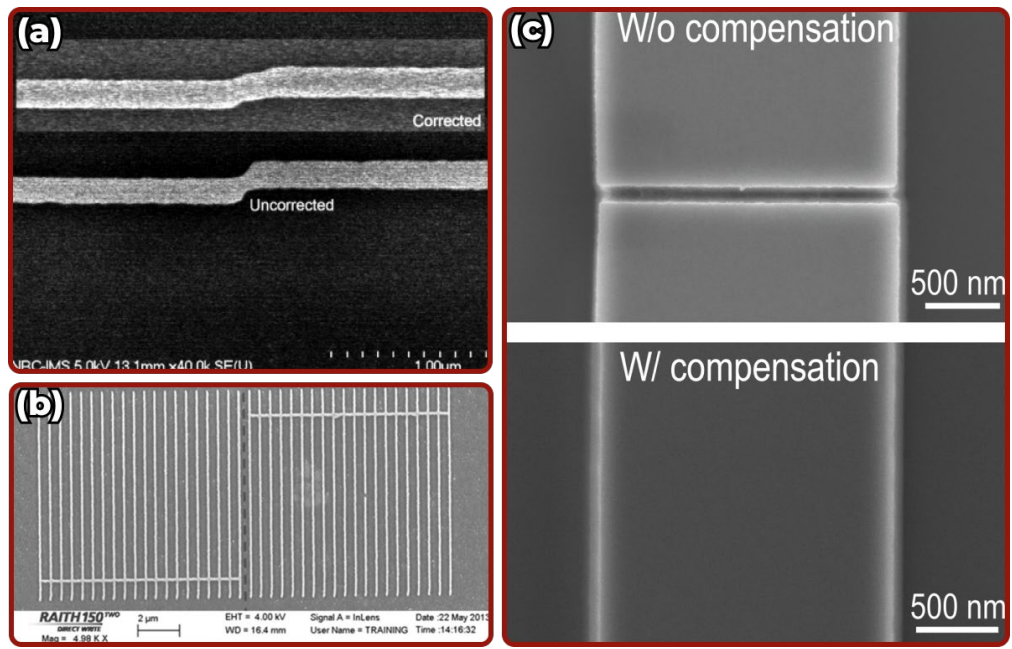

**Figure 5**. Stitching error reduction in e-beam lithography: (a) correction of intentional defect introduced into a silicon photonic wire waveguide, reprinted with permission from [100], Copyright (2012) American Vacuum Society; (b) correction of stitching error with writing field alignment, reprinted with permission from [96], Copyright (2013) American Vacuum Society; (c) stitching error compensation with precisely shifted working fields, reprinted from [104], under a Creative Commons license (https://creativecommons.org/licenses/by/4.0/).

## 2.4. Etching

The next technological step to transfer a design into $Si_3N_4$ layer is wet or dry etching. Each etching process generally consists of three stages: transport of etchants to the wafer surface, reaction of etchants with the surface, and removal of the product species. For both etching processes, the main characteristics include the material etching rate, etching uniformity, and selectivity (the ratio of the resist etching rate to the material etching rate). In addition, special attention should be paid to etching anisotropy and sidewall roughness in the fabrication of PICs, as these factors determine the waveguide surface shape and smoothness, which have a direct impact on scattering losses. For Si, $SiO_2$, and $Si_3N_4$ wet etching, liquid chemicals such as $H_3PO_4$, buffered oxide etch (BOE), and KOH are commonly used [106–108]. Wet etching is a pure chemical process that has serious drawbacks, such as: lack of anisotropy, insufficient process control, and excessive particle contamination. However, wet etching is highly selective, and does not damage the substrate [109,110].

In contrary to wet etching, control of plasma etching process is considerably easier. Furthermore, the plasma processes are less sensitive to external conditions, i.e., wafer temperature. Among the other advantages of plasma etching, there is a higher anisotropy and far fewer particles in the plasma environment compared with liquid media. As mentioned previously, plasma etching is more reproducible than wet etching. In the plasma process, the film surface is etched by an incident flux of ions, radicals, electrons, and neutrals. Although the neutral flux is the largest, most physical damage is related to the ion flux, and chemical damage depends on both the ion flux and radical flux.

Reactive ion etching (RIE), which exploits the synergy of ion and reactive etching mechanisms, is commonly used in the fabrication of $Si_3N_4$ integrated circuits. To etch the small structures and thicker silicon nitride layers, reactive ion etching with an inductively coupled plasma source can be introduced, as it provides a high density of ions and reactive particles, enhancing the etching rate and uniformity [111–114]. Fluorine-based plasmas ($CF_4$, $CHF_3$, and $SF_6$) combined with pure oxygen, nitrogen, or argon are commonly used for selective and anisotropic $Si_3N_4$ etching [112,115,116]. A detailed discussion of the etching processes is beyond the scope of this review. Moreover, a variety of process parameters need to be individually optimized depending on the etching tools used, making it difficult to draw specific dependencies to obtain a low sidewall roughness with high etching selectivity and a vertical waveguide profile. However, certain common assumptions can be identified based on recent studies (**Figure 6**). For low scattering losses, attention should be paid to the balance between ion bombardment, reactive isotropic etching, and surface passivation with fluorocarbon thin film. Higher plasma source power tends to provide higher bombardment, resulting in strong damage to resist and waveguide sidewalls, whereas a high passivation rate is observed at low power values, owing to peculiarities in $CF_4$ or $CHF_3$ based plasma etching [50]. The tailoring of the gas chemistry composition with respect to the process pressure could provide domination of the reactive etching mechanism over ion bombardment, increasing the selectivity and lowering the sidewall roughness [99]. In addition, a cycling process can be introduced to enhance control over the etching and passivation steps [116].

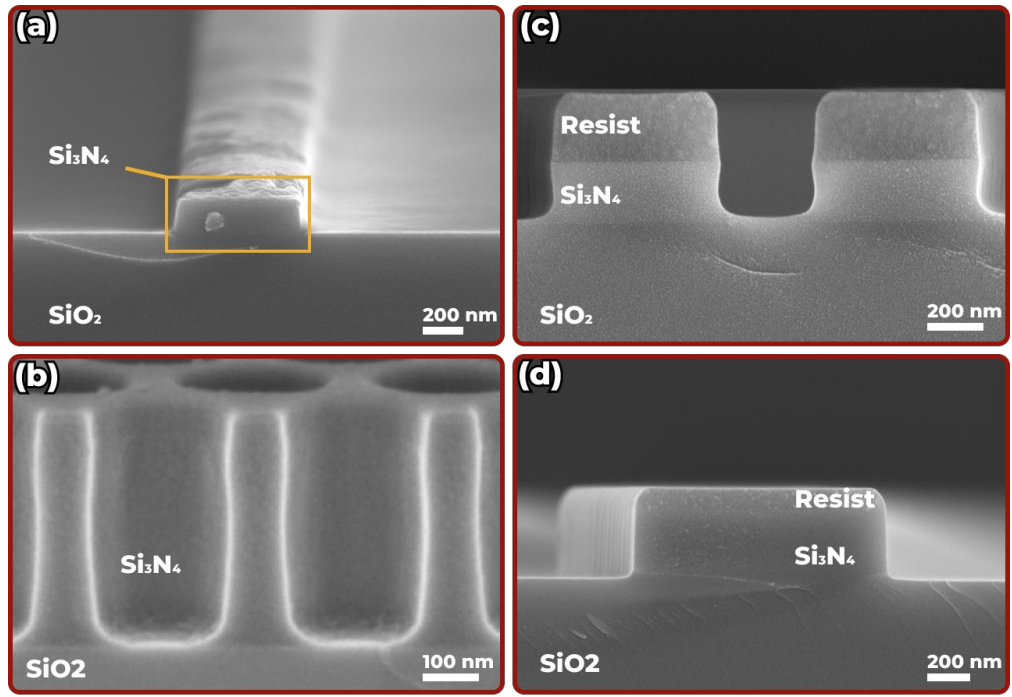

**Figure 6**. Silicon nitride etching: (a) with high plasma source power, leading to high surface roughness; (b) with cycling steps of CHF$_3$ and O$_2$, leading to vertical sidewalls, reprinted with permission from [116], Copyright (2010) American Vacuum Society; (c) with high process pressure, leading to isotropic profile; (d) with optimized process parameters, leading to low surface roughness and vertical sidewalls.

## 2.5. Top oxide synthesis

Most PICs applications require encapsulation of the patterned silicon nitride structures with a high-quality cladding layer to protect them from the environment, to prevent additional losses, or for further functional layer deposition [117,118]. Typically, a silicon dioxide upper cladding layer is used for these purposes (**Figure 7**). A wide variety of techniques for the SiO$_2$ synthesis have been reported [14]: low-temperature oxide (LTO) [39], borophosphosilicate glass (BPSG) [119], SiH$_4$-based PECVD SiO$_2$ [63], TEOS-based PECVD SiO$_2$ [120], deuterated PECVD SiO$_2$ [121], high-temperature oxide (HTO) [50], TEOS-based LPCVD SiO$_2$ [122], sputtered SiO$_2$ [123] and atomic layer deposition (ALD) of SiO$_2$ [124]. A detailed study of each method is beyond the scope of this review. Thus, we briefly introduce key points regarding only common processes in PICs fabrication.

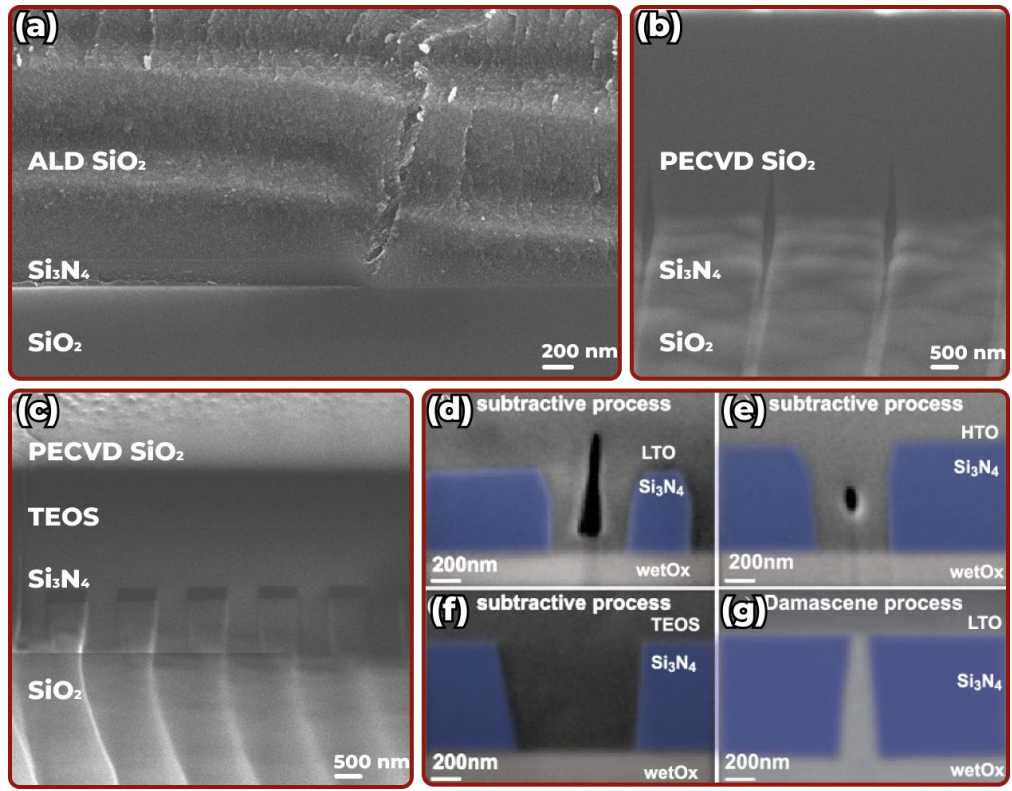

**Figure 7**. Scanning electron micrograph of core-cladding interface for different cladding deposition processes: (a) atomic layer deposition of $SiO_2$ on $Si_3N_4$ waveguide; (b) void formation in grating coupler by $SiH_4$-based PECVD; (c) gap filling in grating coupler by combined TEOS-based LPCVD and $SiH_4$-based PECVD; (d) low-temperature oxide deposition on waveguides realized by subtractive process; (e) high-temperature oxide deposition; (f) TEOS-based LPCVD; (g) low-temperature oxide deposition on waveguides realized by the photonic Damascene process. (d)-(g) reprinted with permission from [73], Copyright (2018) IEEE.

Among low-temperature oxides, LTO and doped $SiO_2$ (BPSG) are commonly used for PICs. The usual processing temperature is in range of 400–450°C because of the lower reaction temperature between silane and oxygen molecules. However, for this reaction, the gas concentration determines the deposition rate. The reagent concentration varies during the deposition process; therefore, it is difficult to create the same conditions for deposition inside the entire chamber. The quality of these films is similar to the $SiH_4$ based PECVD $SiO_2$. High hydrogen concentration and low conformity with extended deposition time make the LTO process less suitable for PICs fabrication. Silicon dioxide doping with phosphorus (adding $PH_3$) and boron (adding $B_2H_6$) could be useful in lowering the oxide glass temperature from 1250 °C to 500–600 °C, which could become a relevant approach in the photonic Damascene process [119]. However, the doped $SiO_2$ films have additional disadvantages: 1)

BPSG films absorb moisture over time (several days), which could further alter the film properties; 2) annealing at high temperature could create defects on the film surface; 3) $P_2O_5$ and $B_2O_3$ clusters are formed on the surface during cooling after annealing.

In contrast, HTO deposition is performed at higher temperatures (typically 800–900 °C) and uses DCS and $N_2O$ as precursors. The surface quality and optical properties of such films are close to those of the thermal oxide, making them the basic technology in qualitative upper cladding layer synthesis in PICs fabrication. However, chlorine contamination and insufficient conformity for the complete coverage of gaps between the high-aspect-ratio structures are the main disadvantages of this technology [39]. TEOS-based processes (LPCVD and PECVD methods) can be introduced to achieve higher conformity and homogeneity for a silicon dioxide film [120,122].

The basic aspects of the PECVD process are indicated in Section 2.1. Silane-based deposition provides fast deposition with the required thickness of several micrometers; however, it results in lower optical and surface quality and high hydrogen content. Moreover, commonly used $SiH_4/O_2$ or $SiH_4/N_2O$ gas mixtures are not desirable because of the pyrophoric nature of silane and nitrogen contamination by $N_2O$. Organosilanes such as tetraethoxysilane (TEOS) have been used since the 1960s to deposit $SiO_2$ using the PECVD method. In addition, TEOS contains silicon and oxygen in a single molecule, reduces toxicity and hazards compared with conventional sources. Furthermore, TEOS-based processes provide films with better conformity and void-free films with superior step coverage, even for small gaps, and retain good mechanical and electrical properties. Although the $SiO_2$ films deposited from pure TEOS exhibit hydrocarbon incorporation, the addition of an oxidant such as $O_2$ or $O_3$ eliminates film contamination. Recently, the introduction of deuterated silane in the PECVD process was proposed for the anneal-free fabrication of low-loss waveguides for application in the C and O telecommunication bands [121] because of the absence of deuterium absorption at 1390 nm. Additional research should be conducted to investigate new materials suitable for $Si_3N_4$ photonic integrated circuits.

Referring to the previously considered fabrication approaches for $Si_3N_4$ integrated photonics (subtractive and photonic Damascene), it is worth noting that none of these technologies use only one method for the upper cladding synthesis. Instead, each approach involves a combination of different $SiO_2$ deposition techniques [14,40,51]. High-quality HTO or TEOS-based LPCVD silicon dioxide forms the first layer that provides the achievable conformity, low absorption, and material interface losses. Next, a PECVD $SiO_2$ layer can be deposited, reducing the time required to obtain the desired cladding thickness.

**2.6. High-temperature annealing**

High-temperature annealing (commonly at 1000–1200 °C for several hours) could be considered as the basis for the fabrication of $Si_3N_4$ PICs with ultralow losses, especially for applications in the C, O, and L telecommunication bands [14,125]. Such thermal processing eliminates the absorption peaks due to the Si–H, N–H, and O–H bonds existing in $SiO_2$ and $Si_3N_4$ [120,126]. Annealing of photonic integrated circuits can be performed both in a nitrogen atmosphere [14,40,51] and in a hydrogen or oxygen ambient. Flash $H_2$-based annealing for a few minutes induces morphological modification of previously fabricated structures in the silicon nitride layer. Furthermore, additional high-temperature oxidation can be performed to effectively passivate and encapsulate the waveguide core and prevent the formation of scattering absorption centers via native oxidation of the $Si_3N_4$ surface [33]. Additionally, dry oxidation can be performed to increase the upper oxide quality before the final high-temperature annealing [127].

A long extremely high-temperature annealing of patterned silicone dioxide is a key process in the photonic Damascene fabrication route. During annealing, wafers are heated slightly above the preform glass transition temperature ($T_G$). In the case of a wet thermally grown $SiO_2$ preform, this temperature corresponds to approximately 1200°C [128]. As the annealing temperature is only slightly higher than that of $T_G$, a prolonged treatment period is required. This provides control over the reflow process and limits the size variations in the waveguide cross-section. Rounding of the waveguide and a sidewall angle of 8° are observed after the reflow; however, a better interface between $SiO_2$ and further deposited $Si_3N_4$ is formed, which provides a smooth waveguide mode transition. Surface-tension-driven smoothing reduces the high spatial frequency components of the sidewall roughness induced by dry etching, resulting in a root mean square roughness lower than 0.2 nm. With respect to the photonic Damascene process silicon nitride chemical mechanical planarization and polishing step, the surface roughness effect on propagation losses is completely eliminated and paves the way to the absorption-limited level of losses [46].

## 2.7. Fabrication of high-efficiency light coupling structures

PICs require an effective coupling mechanism to launch light into a waveguide from a fiber or fiber array unit. Additional losses result from significant mismatch between fiber and waveguide mode field diameters (MFD). To this end, grating couplers and edge couplers have been widely adopted [129–131]. Both coupling strategies have their own benefits and challenges.

Grating couplers (GC) are used for out-of-plane coupling. Such coupling requires fulfilling of phase matching condition or Bragg condition, corresponding to relationship between wave-vector of incident light from fiber and propagation constant of the coupled beam into waveguide [132]. The main optimized parameters for GCs are grating period, core etch depth, grating tooths width and out-coupling angle. Even being carefully optimized, standard fabrication techniques described in Section 2 provide a rather high coupling loss of ~5–7 dB [132]. Additionally, grating couplers with standard design inherently suffer from a limited bandwidth on the order of tens of nanometers and high selectivity to polarization of the input light [130,131]. Recently, several technological approaches for increasing grating coupling efficiency have been proposed, including mirror or DBR deposition [57,133,134], SOI/$Si_3N_4$ integration [135,136], double $Si_3N_4$/$Si_3N_4$ or $Si_3N_4$/a-Si layer gratings [137–139], and complex structure design [140,141], reaching coupling loss as low as 0,6 dB in the C-band (**Figure 8**). The same strategies can be used for grating couplers in the visible spectrum, however, latest achievements in fabrication yield loss higher than 7 dB [142]. It is worth noting, that described approaches lead to increased complexity of PICs fabrication, which currently limits their transition to foundries. Further work should be done for improving deposition of additional layers and developing of more advanced yet simpler constructions.

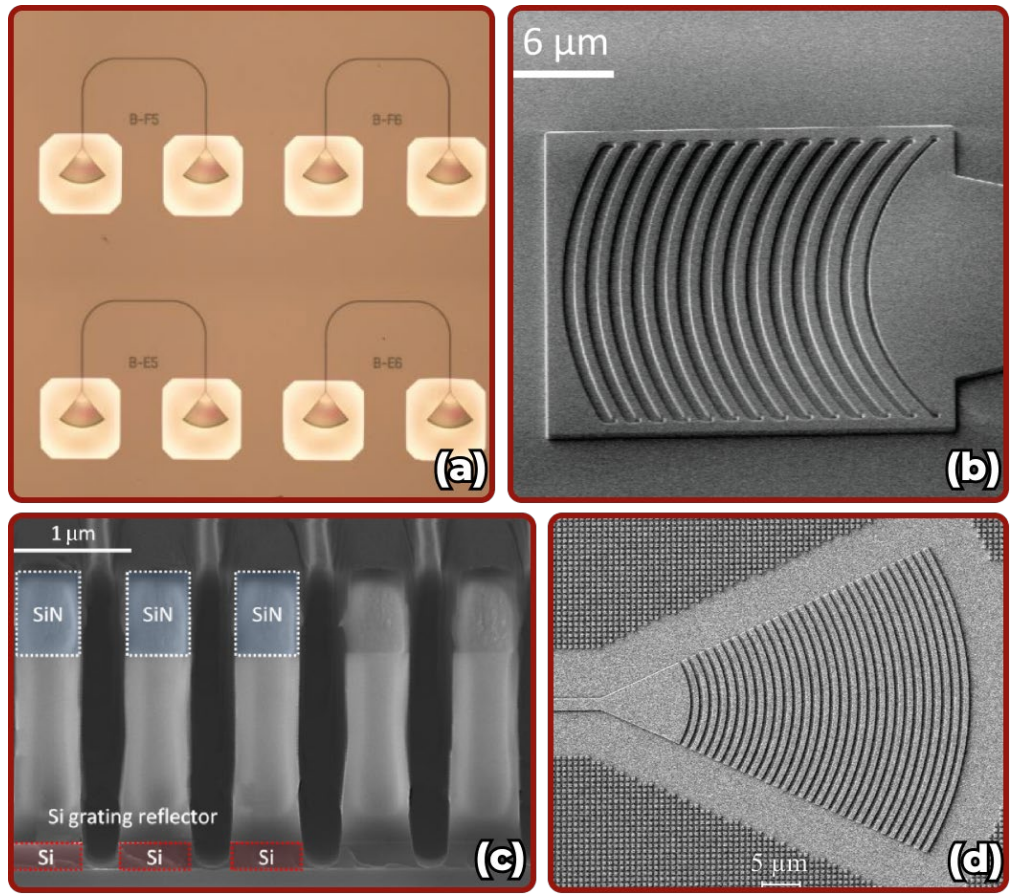

**Figure 8**. Technological approaches for increasing grating coupling efficiency on $Si_3N_4$ photonics platform: (a) array of grating couplers with back-side Al mirror, reprinted from [133], under a Creative Commons license (https://creativecommons.org/licenses/by/4.0/); (b) fabricated $Si_3N_4$-on-SOI dual-level

apodized grating coupler, adapted with permission from [135], Copyright (2014) Optical Society of America; (c) fabricated high-bandwidth SiNx-on-SOI uniform grating coupler, adapted with permission from [136], Copyright (2017) Optical Society of America; (d) fabricated SiNx bilayer apodized elliptical grating coupler with, adapted with permission from [137], Copyright (2018) Optical Society of America.

Compared to grating couplers, edge couplers exhibit superior in-plane light coupling and broader spectral bandwidth if proper match between fiber and waveguide mode field is achieved. However, most edge couplers are designed to entail long adiabatic inversed tapers with lengths exceeding several hundred micrometers, thereby enhancing mode transfer. The lower index contrast of the $Si_3N_4$ waveguide strictly prohibits small bend radii, constraining the implementation of an edge coupler with a small footprint [143]. Moreover, use of common fibers with large MFD require narrow taper tips down to 100 nm, which is of the high complexity for current lithography techniques [130] and posing additional challenges for coupling at visible wavelengths [144,145]. Use of special fibers with lensed facet or ultrahigh numerical aperture tip is a common method for facilitate light coupling, however, complexity of further optical assembly significantly increases, requiring design improvements for efficient coupling with market-available fibers [146]. Recently, edge couplers with high coupling efficiencies have been demonstrated, including double-layer inverted tapers [147,148], inversed-tapered waveguides with fabricated U-groove [149] and oxide gap [150], subwavelength grating-assisted edge couplers [151], and multistage tapers [152,153]. Most of the proposed solutions are realized on SOI photonics platform, however same ideas could be used for $Si_3N_4$ PICs (**Figure 9**).

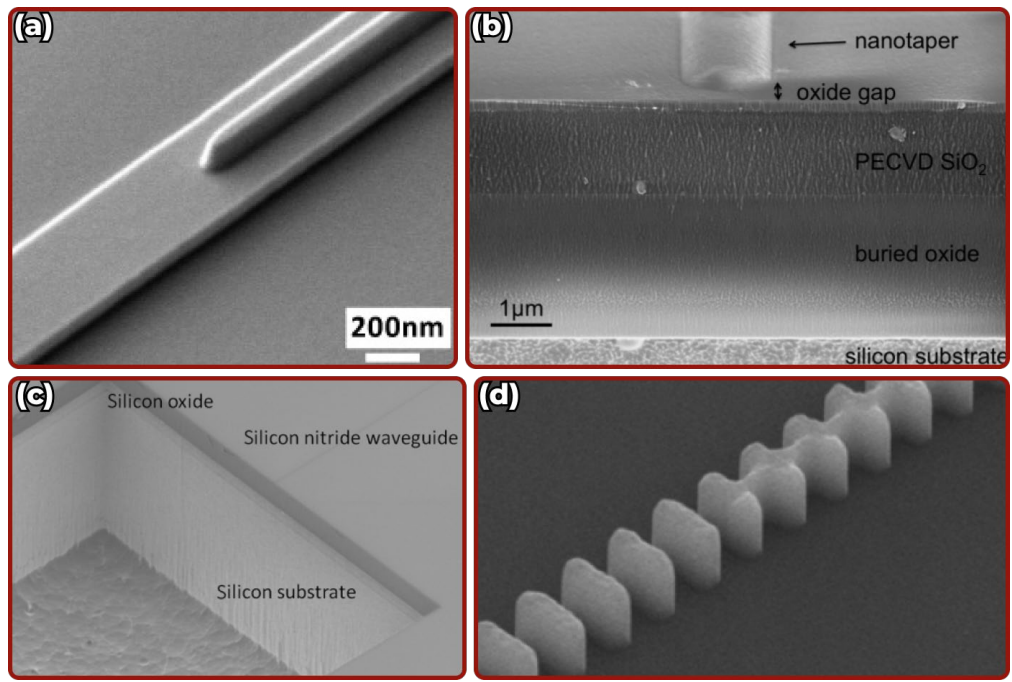

**Figure 9**. Scanning electron micrographs of fabricated edge couplers: (a) suspended dual-level SOI mode size converter, adapted with permission from [147], Copyright (2010) Optical Society of America; (b) inversed tapered silicon waveguide with oxide gap, reprinted with permission from [150], Copyright (2014) IEEE; (c) inversed tapered Si$_3$N$_4$ waveguide with etched U-groove, reprinted from [149], under a Creative Commons license (https://creativecommons.org/licenses/by/4.0/); (d) subwavelength grating-assisted nanotaper, adapted with permission from [151], Copyright (2015) Optical Society of America.

It is worth noting that efficient light coupling through the edge tapers strongly depends not only on the design but also on the fabrication of the optical-grade waveguide facets. Dicing individual chips with dicing machines and laser cutting tools is fast but not the optimal technique because high edge roughness is introduced, leading to high coupling losses. To provide higher coupling efficiency, the edges are typically polished to obtain high-quality optical-grade edges [154]. Although this process is acceptable with separate chips, scaling is difficult in wafer production. To ensure a higher throughput of high-quality optical facets, a plasma etching process was recently proposed [14,155]. To form an optical channel for coupling, a laser lithography step is performed on the front side of the sample where the functional layer is located. The backside laser lithography step could also be performed for PICs with thick core and cladding layers. Here, the main technology figure of merit is the verticality of the resist profile, as it is necessary to avoid the effect of refraction in the process of coupling light from the optical fiber into the optical channel on the chip. Typically, a more than 5 micrometer thick resist layer is deposited on the substrate by spin coating to provide etching of the multilayer $SiO_2/Si_3N_4/SiO_2$ structure and deep reactive ion etching of silicon [14,99,150,156,157].

**2.8. Chemical mechanical polishing**

Chemical mechanical polishing (CMP) process is largely used in the fabrication of electronic and photonic integrated circuits for several applications. One of its primary applications is achieving surface global and local planarity before the fabrication of the multilayer structure. CMP has been widely used to implement the Damascene process, for copper interconnects or the silicon nitride photonic structures fabrication. In the photonic Damascene process, trenches are etched into silicon oxide to create waveguides and other photonic components. After filling the trenches with silicon nitride, CMP is performed to remove excess material [73].

Reducing the top and bottom waveguide surface roughness plays an equally important role in ensuring high quality of the $Si_3N_4$ PICs fabrication [50]. The method to obtain lower losses is using CMP to reduce the upper surface roughness. CMP smooths surfaces using a combination of chemical and mechanical forces. During the CMP process, the polished sample is mounted upside-down in a carrier/spindle, and a slurry is deposited on the pad. Both the pad and the carrier are rotated to remove the material, creating a smooth surface with this approach. There are several factors which influence wafer surface flatness and roughness including slurry composition, pad properties, contact pressure, and process parameters such as wafer/polishing pad rotation speed and polishing time. The most important factors affecting roughness are the pad and slurry parameters. Slurry composition plays a critical role in determining the removal rate and can significantly influence surface roughness. Various abrasive particles (silica, ceria, alumina) are used in the slurry. The particle size, concentration, and chemical reactivity of the slurry affect the material removal rate and the resulting surface roughness [158,159]. The CMP pad properties, such as hardness, elasticity and surface texture also influence flatness and surface roughness. Softer pads can enhance the material removal rate and reduce surface defects but can result in non-uniform surface [160,161]. Choosing the right type of pad and optimizing the pressure helps ensure uniform removal of material across the wafer surface and achieve low roughness. The root mean square roughness could be decreased from 0.38–1.36 nm (before polishing) down to 0.03–0.08 nm (after polishing), which was demonstrated with atomic force microscopy over 5 μm × 5 μm area [14,50,68,81]. This corresponds to propagation losses over 1 dB/m in the C-band. The bottom and top $SiO_2$ cladding layers could also be polished further to achieve even lower losses at the absorption-limited level.

## 3. State-of-the-art si₃n₄-based photonic devices

Silicon nitride integrated photonics was introduced in the early 1990s starting from demonstration of separate photonic elements and, nowadays, it evolved to a mature platform ahead of the other materials (SOI, thin film lithium niobate on insulator (LNOI), III-V waveguides) in terms of losses and scalability [162]. Considering the recent progress in hybrid and heterogeneous integration of active components with passive $Si_3N_4$ PICs [13], it could be used in the manufacture of market-ready devices for applications over broadband spectrum. Evolution path in propagation losses of the $Si_3N_4$ platform is shown on **Figure 10**.

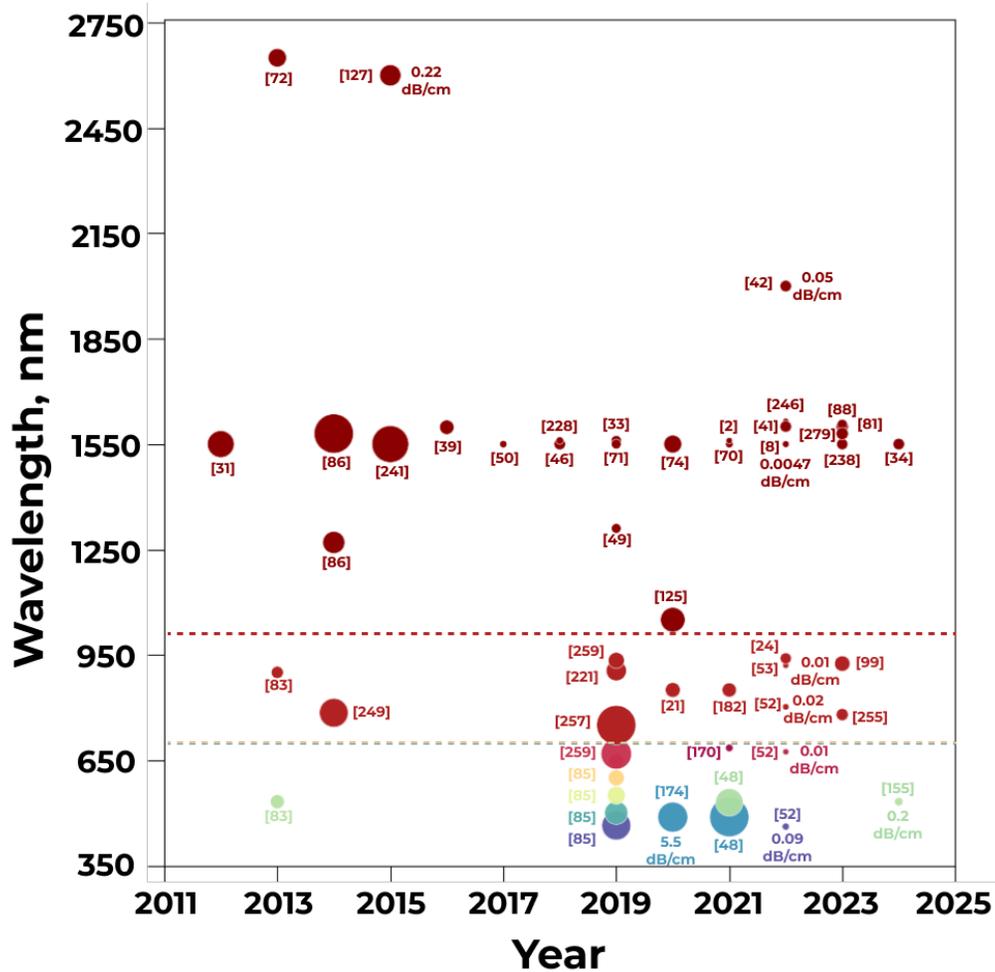

**Figure 10**. Silicon nitride photonics performance evolution in terms of propagation losses (diameter of the circles shows the level of losses and is based on the data from articles, presented in this review; color of the circles represents wavelength).

In the SWIR range, especially within the infrared telecommunication bands (mainly, the O-, C-, and L-band), a broad selection of turnkey photonic devices is available

[Figure 11(a)–(g)], including high-Q microresonators for laser stabilization with sub-MHz linewidth [2,8]; coherent laser ranging systems with wide FOV optical phased arrays [11,12,163,164]; chip-based OCT systems with tunable delay lines [49,105]; ultralow-loss, dispersion engineered circuits for parametric amplification, broadband wavelength conversion, and frequency comb generation [9,10,41,50,70]; reconfigurable photonic circuits for quantum processors, neuromorphic computing and on-chip high-resolution spectrometers [7,165,166]; integrated single-photon detectors with high internal efficiency [167,168]; high-efficient piezoelectric phase shifters [65] and high-bandwidth integrated electro-optic modulators [169] paving the way for technologies to be transferred to other wavelengths.

The recent progress of PICs in the visible and near-infrared ranges is also remarkable (**Figure 11**). Highly coherent integrated lasers [25,170,171], integrated photodetectors [172] with devices capable of single-photon detection [173], micrometer-scale phase shifters and modulators [48], high-FOV integrated optical beam scanning devices [174,175], devices for optogenetics and diagnostic techniques in life sciences [85,176–178], and quantum photonic processors with integrated single-photon laser sources [24,53] have already been demonstrated. In this section, we summarize progress in photonic devices operating in the visible, near-infrared, short-wavelength infrared and mid-infrared ranges with representative works and discuss the prospects and options for further improvement.

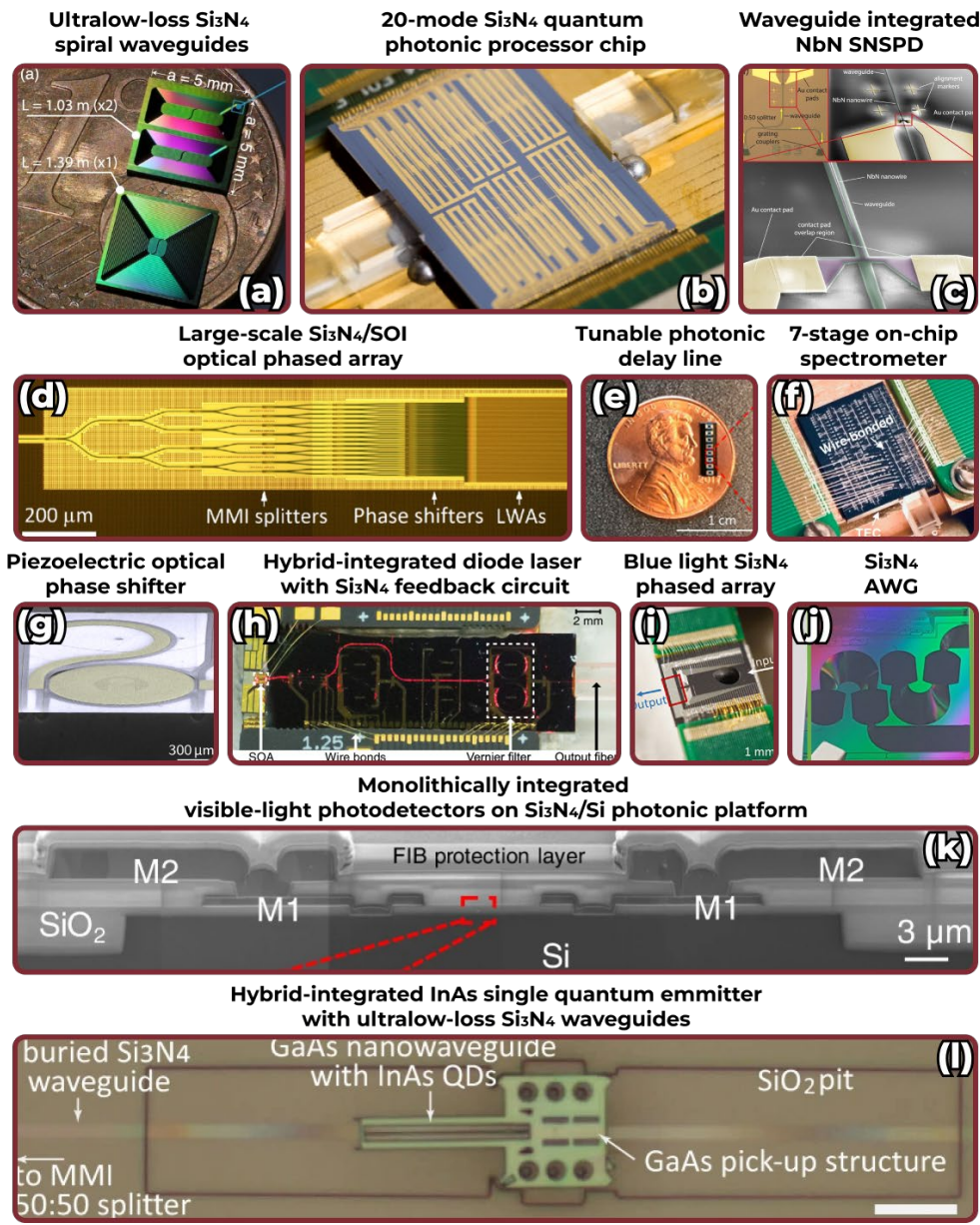

**Figure 11**. Overview of silicon nitride photonic devices for a broad spectrum of applications: (a) ultralow-loss $Si_3N_4$ waveguides, reprinted from [9], under a Creative Commons license (https://creativecommons.org/licenses/by/4.0/); (b) 20-mode $Si_3N_4$ quantum photonic chip, reprinted from [7], under a Creative Commons license (https://creativecommons.org/licenses/by/4.0/); (c) waveguide integrated NbN superconducting nanowire single-photon detector, reprinted from [167], under a Creative Commons license (https://creativecommons.org/licenses/by/4.0/); (d) large-scale $Si_3N_4$/SOI optical phased array, reprinted with permission from [163], Copyright (2020) The Japan Society of Applied Physics; (e) on-chip tunable photonic delay line, reprinted from [105], under a Creative Commons license (https://creativecommons.org/licenses/by/4.0/); (f) broadband on-chip spectrometer,

reprinted from [166], under a Creative Commons license (https://creativecommons.org/licenses/by/4.0/); (g) piezoelectrically tuned optical phase shifter, adapted with permission from [65], Copyright (2018) Optical Society of America; (h) hybrid-integrated visible spectral range diode laser with $Si_3N_4$ feedback circuit, adapted with permission from [170], Copyright (2021) Optical Society of America; (i) blue light $Si_3N_4$ optical phased array, adapted with permission from [174], Copyright (2020) Optical Society of America; (j) $Si_3N_4$ arrayed waveguide grating, reprinted from [178], under a Creative Commons license (https://creativecommons.org/licenses/by/4.0/); (k) Monolithically integrated visible-light photodetector on $Si_3N_4$/Si photonic platform, reprinted from [172], under a Creative Commons license (https://creativecommons.org/licenses/by/4.0/); (l) hybrid-integrated InAs single-photon source with ultralow-loss $Si_3N_4$ waveguides, reprinted from [53], under a Creative Commons license (https://creativecommons.org/licenses/by/4.0/).

## 3.1. Critical building blocks for silicon nitride photonics

It is worth noting that increased complexity and scaling of photonic integrated devices require smooth and low-loss integration of lasers, photodetectors, and modulators with passive $Si_3N_4$ photonic circuits. It pushes scientific community developing next generation integration technologies, novel materials and PICs designs for cost reduction and improved performance. Here, we briefly describe recent advances in lasers, photodetectors, and modulators for $Si_3N_4$ photonics.

### 3.1.1. Highly coherent integrated lasers with ultranarrow linewidth

With growing demand for high throughput datacenters, portable biochemical systems, high-performance and quantum computing, the development of highly coherent on-chip lasers with narrow emission linewidth and ultralow high frequency phase noise level is driving faster [179]. Modern integrated photonics could benefit from development of silicon-based lasers. Silicon, as indirect bandgap material, has limited light emission efficiency, making laser sources on silicon to be the challenging component [180]. Hybrid or heterogeneous integration of group-IV and III-V materials is a classic and mature technology for fabrication of on-chip lasers; however, overcoming the limitations in temporal and spatial coherence still the main challenge [181,182]. The most promising approach to achieve sub-100 kHz linewidth in low-cost and scalable way is based on a semiconductor laser integration with a high-Q silicon nitride microresonator, which is using a self-injection locking scheme. This approach is well developed, having its origins in high-finesse Fabry-Perot (FP) cavities [183,184] and high-Q whispering gallery mode (WGM) microresonators [185,186]. During Rayleigh scattering on internal inhomogeneities and surface roughness, a fraction of the incoming radiation in resonance with the frequency of the selected resonator is reflected back to the laser, providing high optical feedback [187].

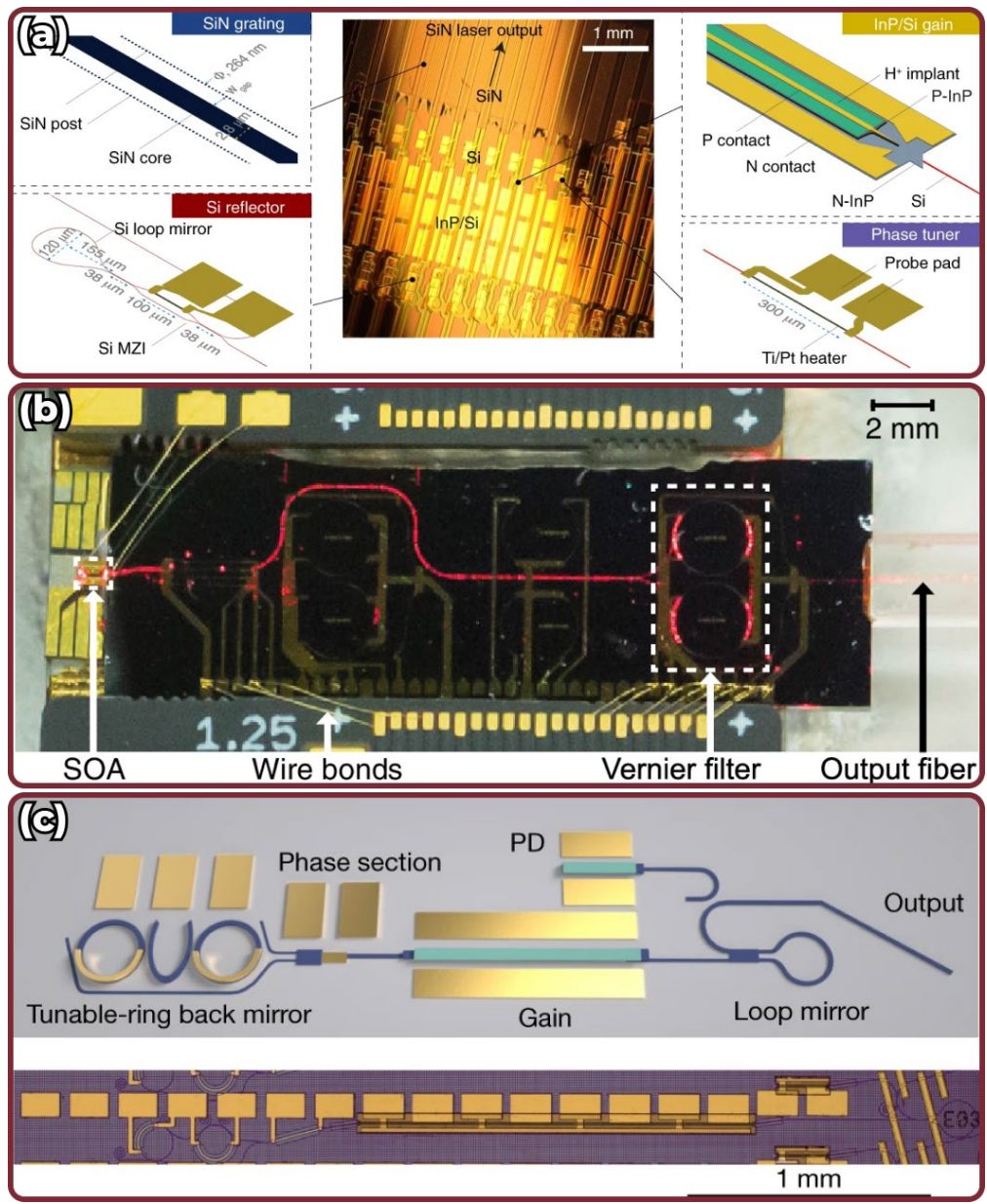

**Figure 12**. Overview of recent achievements in highly coherent integrated lasers: (a) fabricated InP/Si/ Si$_3$N$_4$ DBR laser coupled to Si$_3$N$_4$ high-Q microresonator, reprinted from [93], under a Creative Commons license (https://creativecommons.org/licenses/by/4.0/); (b) Visible spectral range laser coupled to a feedback circuit with a sequence of microresonators, adapted with permission from [170], Copyright (2021) Optical Society of America; (c) Submicrometer dual-ring tunable laser, reprinted from [171], under a Creative Commons license (https://creativecommons.org/licenses/by/4.0/).

Various integration techniques and microresonator circuit designs have been proposed to overcome technological constraints, especially in the VIS and NIR wavelength ranges (**Figure 12**) [25,54,93,170,171]. Coupling the commercial distributed feedback laser to an ultrahigh quality factor SiN microresonator (TriPleX platform, Q ~ 220 millions) delivers 1–3 Hz linewidth and frequency noise of 0.2–1 $Hz^2$/Hz in the C-band [54,93]. A complex optical feedback circuits design with additional loop-mirrors [25] and multiple microresonator configurations [170] allows effective linewidth reduction down to ~ 1.15 MHz in the near-UV-VIS range (410–532 nm) and to ~2.3 kHz at 685 nm with very high potential to improve it with further technology optimization. Additional microheaters above a ring resonators section allows wide tunability with ~ 2.8 kHz linewidth across 20 nm tuning range in the NIR wavelength band [171].

*3.1.2. Broadband and high-efficiency integrated photodetectors*

Apart from the integrated laser sources, future integrated photonic devices require high-efficiency integrated photodetectors (PDs). Modern PDs should meet such requirements as high efficiency, large bandwidth, low dark current and maximum output power, being compatible with photonic integrated circuits fabrication (mostly, with CMOS pilot line). SWIR photodetectors became a mature technology with hybrid and heterogeneous integration of III–V devices with $Si_3N_4$ photonics (**Figure 13**). Group-IV and III–V materials photodetectors are commonly used for the detection of optical signals with high responsivities up to 96%, less than 20 nA dark current and high bandwidths over 30 GHz [188–190]. For an on-chip detection at visible and NIR wavelengths, the integration of Si PDs has the potential to provide high-efficiency detection due to enhanced absorption of silicon at wavelengths lower than 1100 nm. Optical signal transfer from $Si_3N_4$ waveguides to monolithically or heterogeneously integrated P-I-N junctions in silicon with end-fire coupling or wave leakage is demonstrated [172,191]. High quantum efficiencies up to 60% at visible (30% at NIR) wavelength bands and wide optoelectronic bandwidths have been achieved.

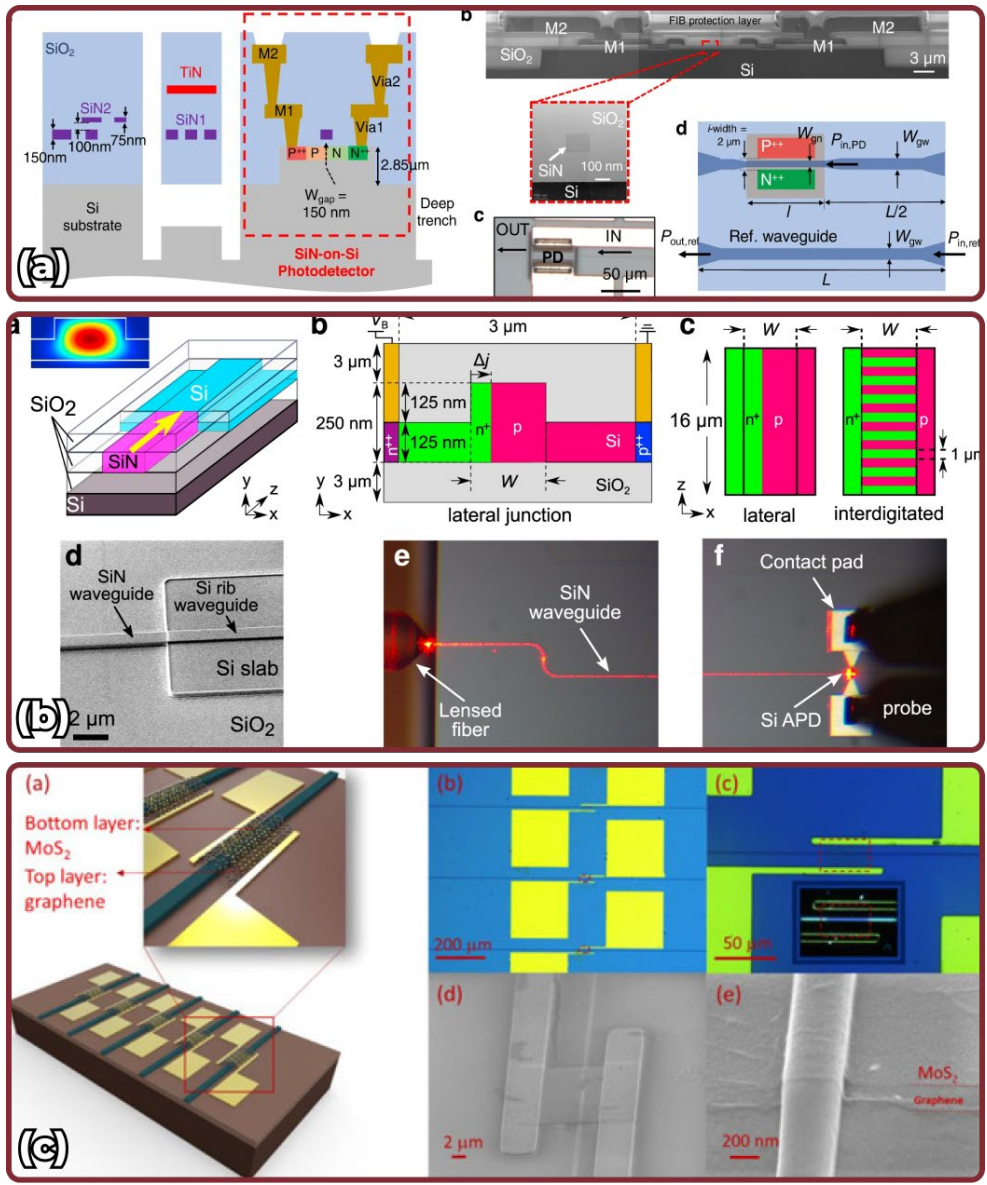

**Figure 13**. Integrated photodetectors on silicon nitride photonic circuits: (a) Visible spectrum integrated photonics platform with leaky-wave bulk Si photodetectors, reprinted from [172], under a Creative Commons license (https://creativecommons.org/licenses/by/4.0/); (b) Integrated silicon-based avalanche photodetector for visible light, reprinted from [191], under a Creative Commons license (https://creativecommons.org/licenses/by/4.0/); (c) $Si_3N_4$ waveguide-integrated graphene/$MoS_2$ heterostructure photodiode for visible light, reprinted with permission from [192], Copyright (2018) WILEY-VCH Verlag GmbH & Co. KGaA, Weinheim.

Apart from III-V and silicon-based devices, two-dimensional (2D) materials, such as graphene [193], platinum deselenide [194], 2D-based heterostructures [192] can bring advantages in on-chip PDs bandwidth and transparency range [195]. Responsivity over 440 mA W$^{-1}$ and bandwidth up to 67 GHz have been recently demonstrated with a low-loss silicon nitride platform at visible and telecom wavelengths. Despite considerable achievements in 2D material technologies, much more efforts are required to incorporate them into silicon or silicon nitride platforms. This includes defects and contaminations elimination [196], prevention of waveguide damage during transfer [192], and design optimization for efficient heat dissipation in photodetector arrays [197].

*3.1.3. Micrometer-scale phase modulators*

Various applications including optical computing, routing, and switching require integrated phase shifters and modulators with low-loss, high bandwidth, and high-power efficiency. In Si$_3$N$_4$ photonics thermo-optical control has been the main approach for light modulation over 30 years, offering simple and relatively scalable technology with kHz-speed limit [198] (**Figure 14**). However, it is quite challenging to address low power consumption and footprint requirements with it due to low thermo-optic coefficient, which results in tens of milliwatts for π-phase shift (P$_\pi$). Convenient, millimeter-long phase shifters based on metal films (Ti [99], Pt [199], or metal stacks [200]) require significant design optimization for thermal crosstalk mitigation and higher modulation speeds [201,202]. Phase shifters design exploiting folded Si$_3$N$_4$ waveguide with air trenches allows fabricating a power-efficient structure with low insertion loss over 1.9 dB and ultralow heating power (for P$_\pi$ shift) of 0.78 mW at 445 nm [203]. An alternative solution with adiabatic microring resonators operating in a strongly over-coupled regime in blue (488 nm) and green (530 nm) wavelengths is proposed in [48]. High power efficiency and over 300 kHz modulation speeds, which requires only 0.68 mW heating power for P$_\pi$ shift with less than 0.9 dB amplitude variation, were demonstrated.

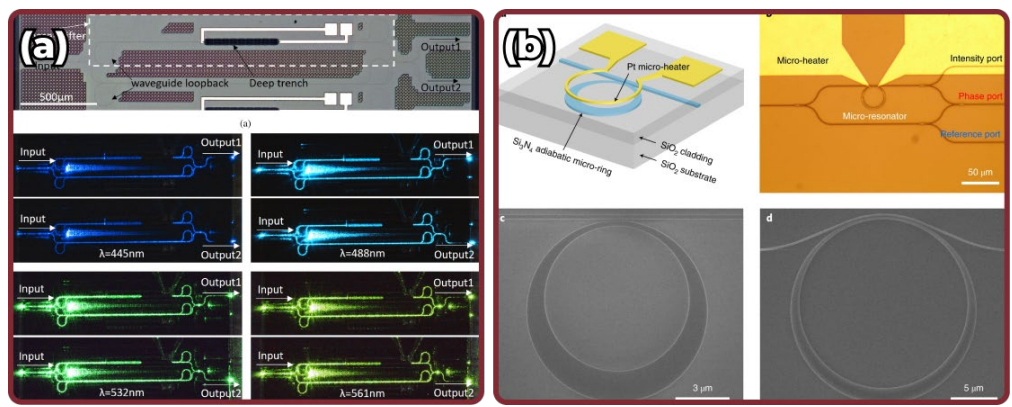

**Figure 14**. Advances in thermo-optic phase shifters for silicon nitride photonic circuits: (a) Phase shifters for visible light with a suspended heater structure, adapted with permission from [203], Copyright (2022) Optica Publishing Group; (b) Visible-spectrum phase modulation in a strongly over-coupled Si$_3$N$_4$ microresonators,

Fast optical modulators are the key components of all the photonic integrated platforms (**Figure 15**). Approaches based on piezoelectric light modulation on $Si_3N_4$ PICs are recently proposed [65,204]. In contrast to typical thermo-optic approaches, PZT-based piezoelectric actuators demonstrate greater performance, resulting in a tuning response up to 1 MHz with $V_\pi L_\alpha$ only 1.1 V dB and low propagation losses of 0.3 dB/cm [65]. It is worth noting that such an approach can significantly improve practically useful integrated optical networks performance with thousands of devices, as tuning efficiency improves with device shrinkage. However, close attention should be paid to actuator fabrication due to excess waveguide loss. Aluminum nitride is another common material used in integrated piezoelectric actuators. Efficient stress-optic tuning in silicon nitride microresonators with Mo/Al/AlN actuators was demonstrated, achieving a tuning rate of 5.3 MHz with ultralow power consumption (less than 300 nW) and preserving ultralow waveguide loss of 0.02 dB/cm [204]. Further integration with microelectronic integrated circuits and coherent integrated lasers could attain compact tunable soliton microcomb sources for cryogenic operation, coherent solid-state LiDAR systems, and RF photonics [205].

For higher modulation speeds, there is no widespread solution for $Si_3N_4$ platform. Plasma dispersion effect-based modulators, which are used in silicon, are not available for dielectrics. Significant second-harmonic generation has been reported for silicon-rich $Si_3N_4$, but few Pockels effect-based modulators have been demonstrated on $Si_3N_4$ photonic circuits [169,206]. There is a great demand for high-frequency low-loss integrated optical modulators with high modulation efficiency and bandwidth, compatible with ultralow-loss $Si_3N_4$ photonics. However, it remains challenging due to fabrication issues and materials incompatibility. That is why material for integrated modulators is a widespread field of research nowadays. Lithium niobate [207], indium tin oxide [208–210], lead zirconate titanate [169], barium titanate [211], and electro-optical polymers [212,213] are promising platforms for high-efficiency electro-optic modulators; however, only few solutions achieve valuable results.

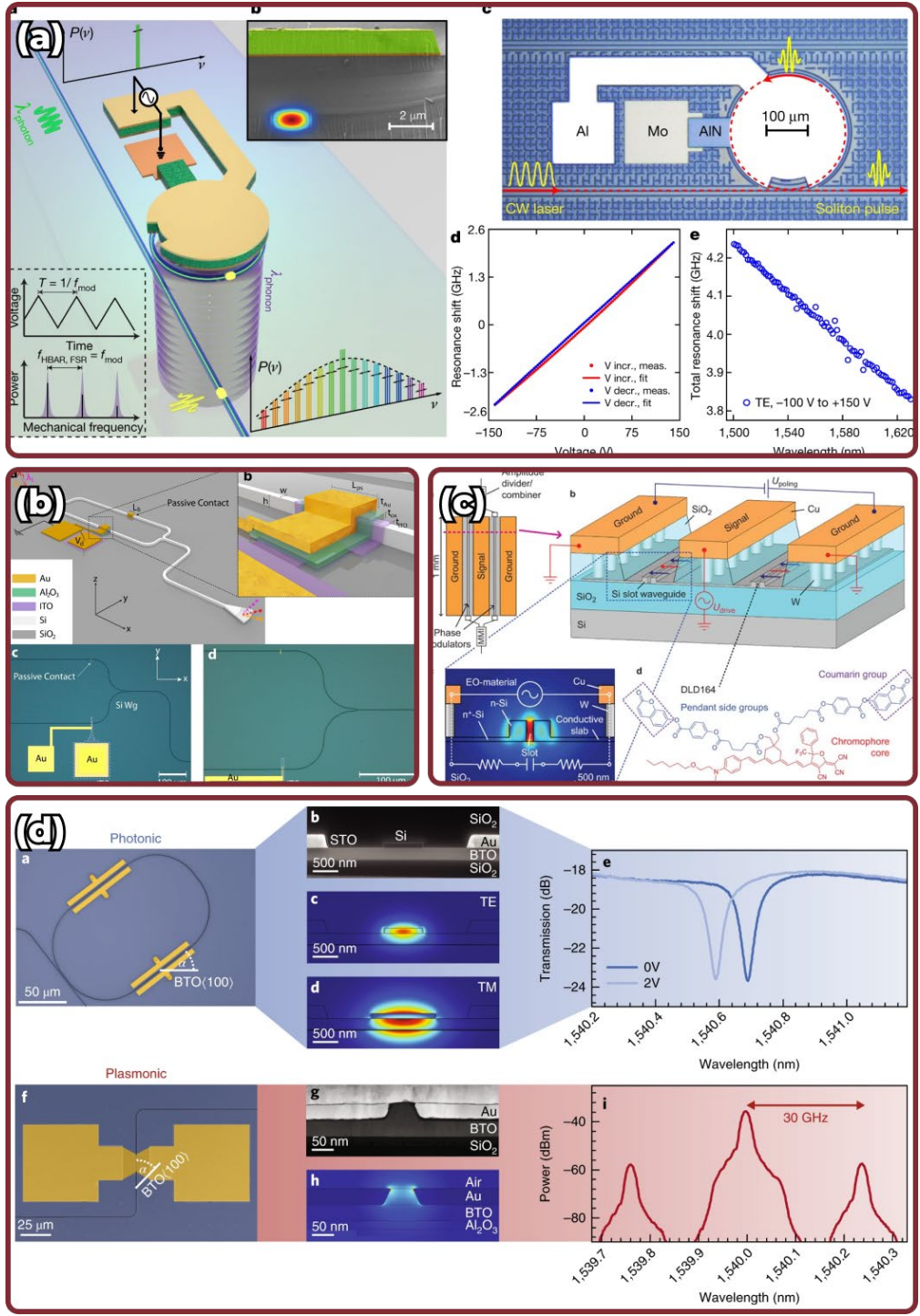

**Figure 15**. High-frequency optical modulators for integrated photonics: (a) Bidirectionally tuned integrated AlN piezoelectric actuator, reprinted with permission from [the Licensor]: [Nature][204], [COPYRIGHT] (2020); (b) Integrated ITO-based

electro-optic modulator on SOI, reprinted from [208], under a Creative Commons license (https://creativecommons.org/licenses/by/4.0/); (c) Silicon-organic hybrid electro-optic modulator, reprinted from [213], under a Creative Commons license (https://creativecommons.org/licenses/by/4.0/); (d) Barium titanate-based integrated electro-optic modulator on SOI, reprinted with permission from [the Licensor]: [Nature][Nature Materials] [211], [COPYRIGHT] (2019).

The proposed electro-optic modulators have demonstrated high data rates operation and high bandwidths (up to 33 GHz) with 1 dB/cm modulator loss and 12 dB total losses between the open and PZT-coated waveguide regions, which are not limited by intrinsic PZT response, but by device design and characterization techniques [169].

### 3.2. Emerging applications of silicon nitride photonics

Silicon nitride photonics is on the upswing due to the broadband nature of the material, allowing applications from visible to mid-infrared wavelengths in datacom, optical signal processing, optical computing, and biophotonics [198]. In this section, we review state-of-the-art silicon nitride photonic devices.

#### 3.2.1. Life sciences applications

Emergence of new viral diseases poses challenges of progressive complexity for modern medicine. Current laboratory diagnostics and drugs development are on intense improvement with application of biophotonic integrated circuits. The main objective is to create portable and precision devices for non-invasive analysis and biosensing applications, e.g., in optogenetics, fluorescent microscopy and optical coherence tomography [214,215].

High-resolution optogenetic stimulation and fluorescence imaging applications require compact circuits that can illuminate small sections of the brain tissue with minimal damage and cover a wide spectrum to effectively excite commonly used opsins. Recently, an implantable neurophotonic probe (**Figure 16a**) for rodent brains optogenetic stimulation was demonstrated as a proof-of-concept for a low-loss visible light photonic platform [85]. Moreover, the benefits of the low-loss silicon nitride visible photonics platform could be used for stimulation of commonly used fluorophores. Miniature and robust multi-color laser engines (Figure 16b) based on TriPleX photonics chips were recently demonstrated, combining four fluorophore excitation wavelengths (405 nm, 488 nm, 561 nm, 640 nm) for the entire visible range coverage [176].

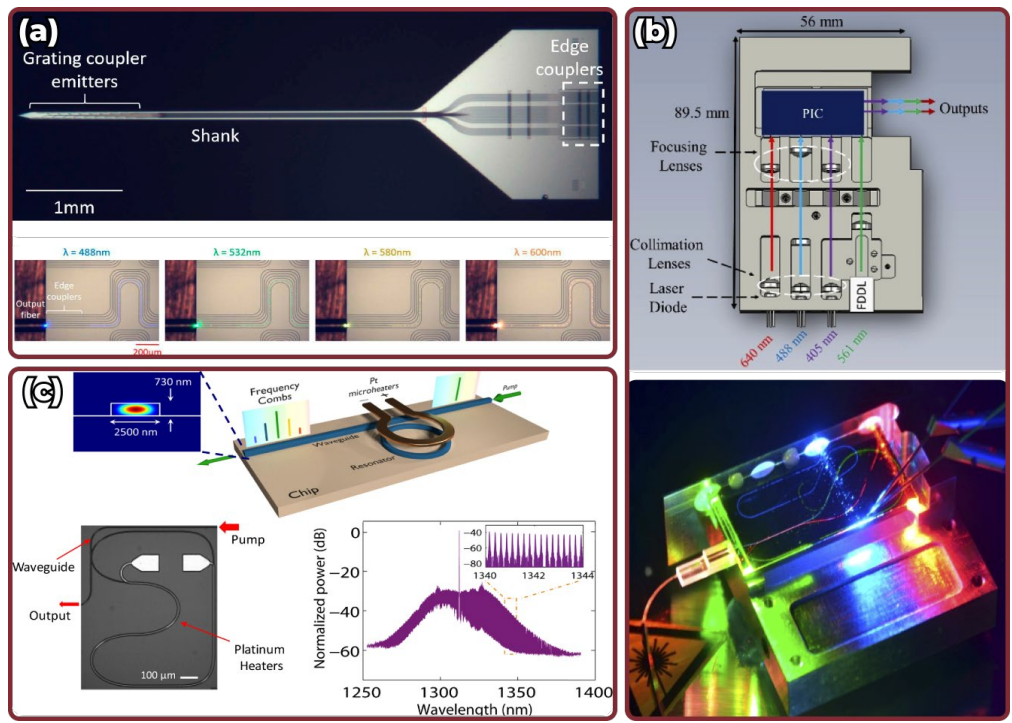

**Figure 16**. Silicon nitride photonics for life sciences: (a) Proof-of-concept implantable neurophotonic probe for optogenetic stimulation of rodent brains on visible light silicon nitride photonics platform, adapted with permission from [85], Copyright (2019) Optical Society of America; (b) Multi-color laser engine device for stimulation of commonly used fluorophores on low-loss TriPleX $Si_3N_4$ platform, adapted with permission from [176], Copyright (2021) Optical Society of America; (c) Chip-scale microresonator-based frequency comb light source for optical coherence tomography, adapted with permission from [49], Copyright (2019) Optical Society of America.

Chip-based optical coherence tomography (OCT) systems could be a powerful solution for low-cost and portable devices that are suited for non-invasive analysis of tissue microstructures in medical applications. Recently, several concepts for on-chip OCT imaging at NIR and SWIR optical bands have been proposed [49,105,177]. Tomograms with clinically accepted contrast and resolution could already be generated [Figure 16(c)], thanks to $Si_3N_4$ broad bandwidth arrayed waveguide gratings (AWG), ultralow-loss meter-long delay lines, and high-power frequency comb sources. However, the development of market-ready devices will require integration of $Si_3N_4$ components with laser sources and photodetectors, further reduction of transmission losses, and design optimization.

Low-loss $Si_3N_4$ photonics also paves the way for developing real-time biosensing devices [216]. Typically, optically passive biosensors architectures utilizing Mach-Zehnder interferometers and high-Q microresonators, using refractive index change as the sensing mechanism. Current state-of-the art detection limit of sensors ranges from $10^{-7}$ to $10^{-4}$ refractive index unites, depending on configuration and target molecule [22,217,218]. To further minimize the demand for expensive tunable lasers and spectrometers, addition of cascades of microring resonators and arrayed waveguide gratings to current sensors schemes was recently proposed [219,220]. Further work should be done to develop techniques for integration of PICs and microfluidic chips for precise and accurate control of liquid volumes, flows and pressures [214,216].

*3.2.2. Solid-state light detection and ranging*

Automotive, industrial automation, environmental and security of critical infrastructure were revolutionized with the development of 3D light detection and ranging (LiDAR) solutions in early 2000s. A new stage of LiDAR sensors development has come with advances in photonic integrated circuits. The main effort is focused on research in high-coherence frequency modulated continuous wave sources and solid-state scanners [201].

The need for rapid and precise light beam scanning in LiDAR sensors is driven by the increasing demands of driving assistance systems and intense research in quantum information processing, biological sensing, and underwater surveying applications. Optical phased array (OPA) technology for beam scanning is attracting considerable attention as it provides high-resolution and wide-angle field of view (FOV) over two dimensions with scalable silicon photonics [221]. Apart from modern mechanical and microelectromechanical scanners, OPA-based beam steerers operate by tuning the phase difference between adjacent channels in transmitter, requiring no moving parts and having high potential to be integrated with on-chip lasers and photodetectors [222]. Being limited by silicon bandgap, further research has been conducted to bring this technology to silicon nitride photonics platform [201]. Beam steering at blue wavelengths is recently demonstrated [174], achieving 50° FOV and 0.17° resolution with sparse spatial arrangement of the 64 aperiodically positioned end-fire emitters (**Figure 17**a).

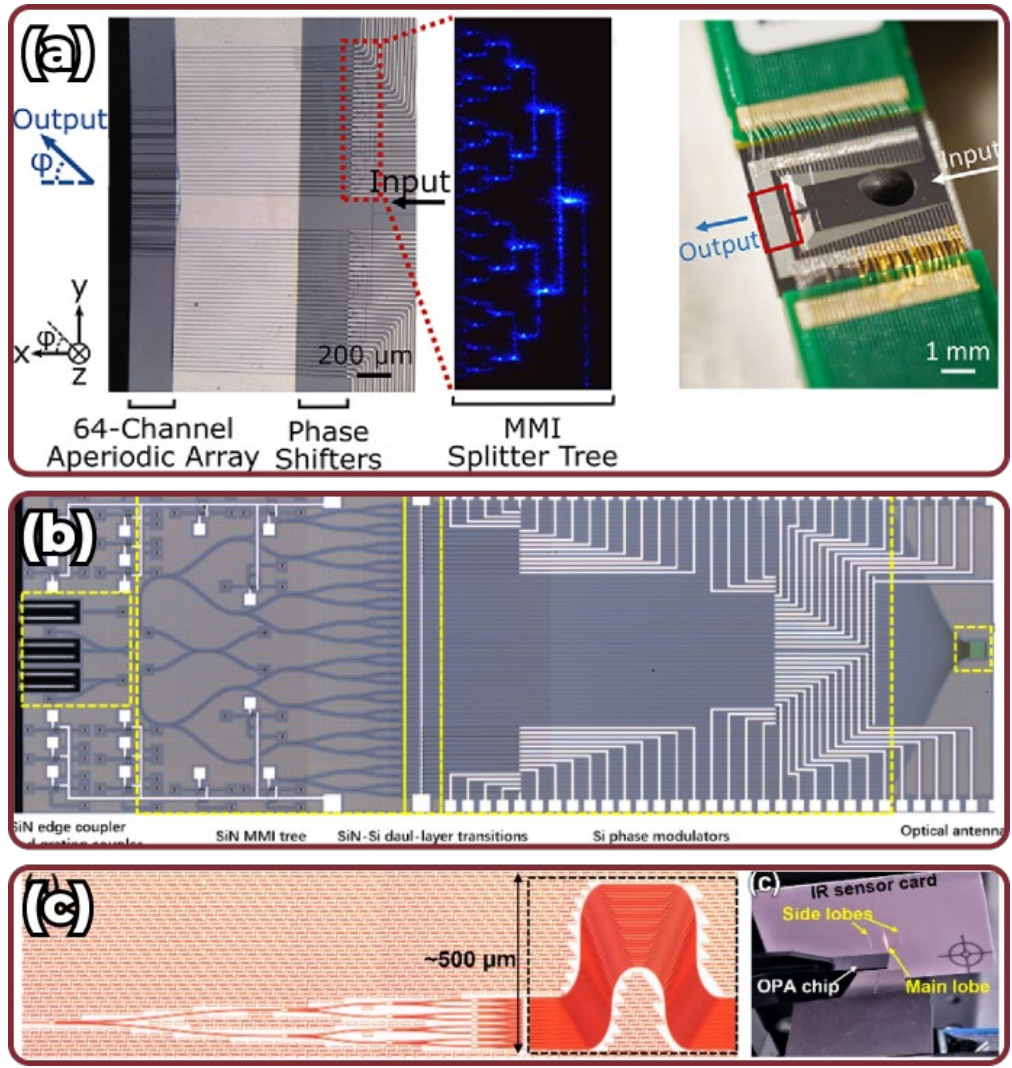

**Figure 17**. Optical beam scanning in silicon nitride photonics: (a) Blue light end-fire optical phased array, adapted with permission from [174], Copyright (2020) Optical Society of America; (b) $Si_3N_4$/Si dual-layer waveguide grating antenna optical phased array, adapted with permission from [4], Copyright (2020) Chinese Laser Press; (c) Passive end-fire optical phased array incorporating arrayed waveguide delay lines, reprinted from [11], under a Creative Commons license (https://creativecommons.org/licenses/by/4.0/).

However, a higher FOV together with high resolution requires scaling up to hundreds of emitters, bringing complexity to photonics net design and control of thermo-optic phase shifters due to lower mode confinement and low thermo-optic coefficient in silicon nitride photonics. Several solutions have been proposed to overcome these limitations, including $Si_3N_4$/Si dual-layer technology (Figure 17b) [4] and passive beam scanning with arrayed waveguide delay lines (Figure 17c) [11]. The combination of silicon and silicon nitride in OPA benefits from low-loss $Si_3N_4$ passive devices and high-efficiency SOI active devices, resulting in a two-dimensional FOV of 96°x14° and 2.3°x3.1° spot size [4]. Constructing the arrayed waveguide delay lines results in wavelength-dependent beam steering over 23.8° without additional electronic circuit [11]. However, further integration of tunable laser diodes could be quite challenging in terms of low phase noise, which is also required for long-distance FMCW laser ranging [223].

Utilizing recent achievements in narrow-linewidth lasers and nonlinear optics, an outstanding cycle of works was recently been performed, showing realization of coherent LiDAR engines for massively parallel detection and ranging [12,164,224–226]. Proposed architecture utilizes ultrahigh quality factor $Si_3N_4$ microresonator for generation of broadband dissipative Kerr solitons (DKS) with high repetition rate. Frequency modulation of pump laser accurately transfers to each sideband of optical microcomb. As a result, more than 5 megapixels per second measurement rates are achieved with cm-level precision ranging, showing possibility for building compact LiDAR-sensors with more than 60 simultaneously working channels [224], however further work should be done to combine such sources with recently developed optical phased array beam steerers with respect to specifications, desired by Level 4 and 5 autonomous vehicles [227].

*3.2.3. Nonlinear applications*

Silicon nitride has emerged as the leading material platform for integrated nonlinear photonics due to ultralow losses, relatively high mode confinement, anomalous group-velocity dispersion possibility, strong Kerr nonlinearity, and the absence of two-photon absorption in the telecom bandwidth [228]. The main research in this field is focused on integrated optical frequency combs (OFC) sources. A low threshold power level at a high repetition rate is required in a diverse number of applications, including metrology and sensing, parallel coherent communications, comb-based light detection and ranging, low-noise microwave signals generation, optical frequency synthesis and calibration of astronomical spectrometers [205]. From 2014, continuous work in understanding microresonator-based optical frequency combs physics and operation regimes is ongoing, reaching several milestones – first demonstration of temporal DKS in crystalline $MgF_2$ microresonators [229], dispersive waves generation [230], octave-spanning DKS frequency combs [231], platicon microcombs [232], excitation of breather solitons [233]. Exploiting the benefits of the developed fabrication and advanced dispersion engineering techniques, the generation of broadband and high repetition rate dissipative Kerr solitons via high-Q microresonators is demonstrated within subtractive, TriPleX, and photonic Damascene processes [10,70,234] with more than 100 GHz free spectral range (**Figure 18**a–c), approaching the characteristics of DKS in crystalline microresonators. However, the viability of current technology could be increased with the development of new approaches for photonic integration and material synthesis [205].

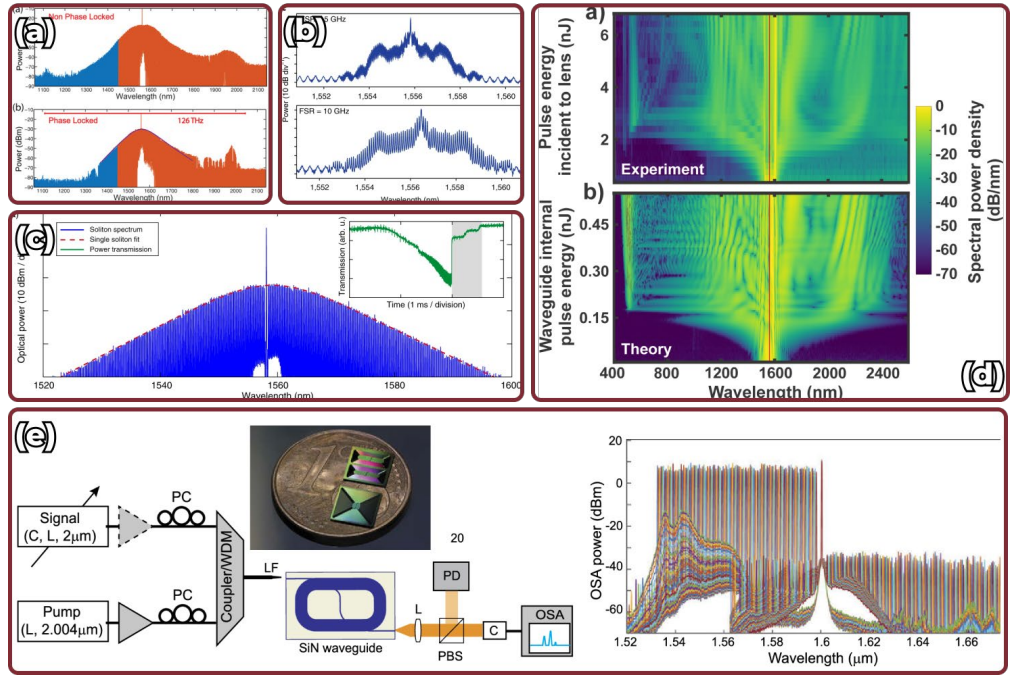

**Figure 18**. Nonlinear optics in silicon nitride photonics: soliton frequency combs generation within (a) subtractive, reprinted with permission from [70], Copyright (2018) WILEY-VCH GmbH; (b) TriPleX, reprinted with permission from [the Licensor]: [Nature][Nature Photonics], [COPYRIGHT] [54]; (c) photonic Damascene fabrication processes, reprinted from [10], under a Creative Commons license (https://creativecommons.org/licenses/by/4.0/); (d) Supercontinuum generation at telecom pump wavelengths, adapted with permission from [235], Copyright (2017) Optical Society of America; (e) Broadband wavelength conversion in the L-band, adapted with permission from [41], Copyright (2022) Optica Publishing Group.

Nonlinear silicon nitride applications are not limited by OFC generation [236]. Broadband and polarization selective wavelength conversion from the O-band to the MWIR wavelength range is demonstrated (Figure 18) in high-confinement $Si_3N_4$ waveguides engineered for zero dispersion at these wavelengths [41,237,238]. Varying silicon nitride optical properties via enrichment with silicon or nitrogen [235,239,240] and careful core geometry engineering results in broadband supercontinuum generation from visible and up to mid-infrared wavelengths [235,241] (Figure 18). Optical power amplification is another nonlinear interaction-based process successfully implemented on silicon nitride photonics. On-chip parametric amplification with a high parametric gain of up to 35 dB and broadband coverage (from NIR to SWIR range) is demonstrated, offering low power consumption and a compact footprint for devices in medical, quantum, and communications applications [242,243]. Moreover, lasing based on stimulated Brillouin [244–246] and Raman scattering [247–249] is demonstrated, allowing for compact systems in atomic, molecular and optical physics experiments, narrowband microwave photonic filters and lasers at wavelengths for which semiconductor-based gain media are not readily available.

*3.2.4. Reconfigurable linear photonic circuits*

The need for higher computing performance and constant increase in data processing volumes require new approaches to the development of large-scale processing circuits [250,251]. Photonic integrate circuits could be involved in architectures of quantum processors as a main processing unit - reconfigurable linear optical circuit [7,53]. Moreover, application of PICs in other architectures was proposed, i.e., trapped ions [246–249] and neutral atoms [252–255]. In this section, we focused attention on application of silicon nitride platform for quantum and neuromorphic processing.

Quantum photonic processors consists of single-photon sources, linear optical circuits and single-photon detectors. Depending on used photon sources type two wavelengths are used. Non-deterministic photons are generated by quantum dots at wavelength in the 900 nm to 970 nm wavelength range [256,257], deterministic photons are generated by parametric conversion in on-chip structures: spirals and resonators and operate at 1550 nm wavelength [258,259].

Recently, significant progress has been achieved in the fabrication of complex devices with integrated single-photon sources. The series of quantum photonic processors (8x8 [5], 12x12 [6], and 20x20 [7] input/output modes) was presented by QuiX for the C-band (**Figure 19**a). The backbone of these processors is a low-loss waveguide network based on a tunable Mach–Zehnder interferometer (MZI). The fabricated devices demonstrated high amplitude fidelities for the permutations and Haar-random transformation up to 99.5% and 97.4%, respectively, which paves the way to fully integrated quantum computers. However, the manufacturing of high-quality quantum dot single-photon sources operating in 1550 nm wavelength range is still a complex problem due to the increased physical dimensions and higher quantum dots exposure to material stress and impurities [260].

In fact, commercially available non-deterministic single-photon sources with higher efficiency operate in the 900 nm to 970 nm wavelength range [261,262]. Therefore, to implement an optical quantum computer, it is necessary to develop a low-loss and scalable PICs platform compatible with this wavelength range. A 12-mode quantum photonic processor with an integrated InAs light source (Figure 19b) that emits light in the NIR wavelength range was recently demonstrated [53]. The photonic processor reconfigurability was demonstrated by implementing 100 Haar-random matrices with measured fidelity up to 0.986, exhibiting a sub-dB propagation loss level per 1 cm of photonic network. Improved fabrication technology [263] allowed the development of wafer-scale technology for the integration of InAs quantum emitter single-photon sources into silicon nitride waveguides (Figure 19c) with propagation losses as low as 0.01 dB/cm at 930 nm and high single-photon Fock-state purity [53]. However, the relatively low single-photon coupling efficiency in the waveguide was demonstrated [264], thus requiring optimization of the device design and quantum dot positioning.

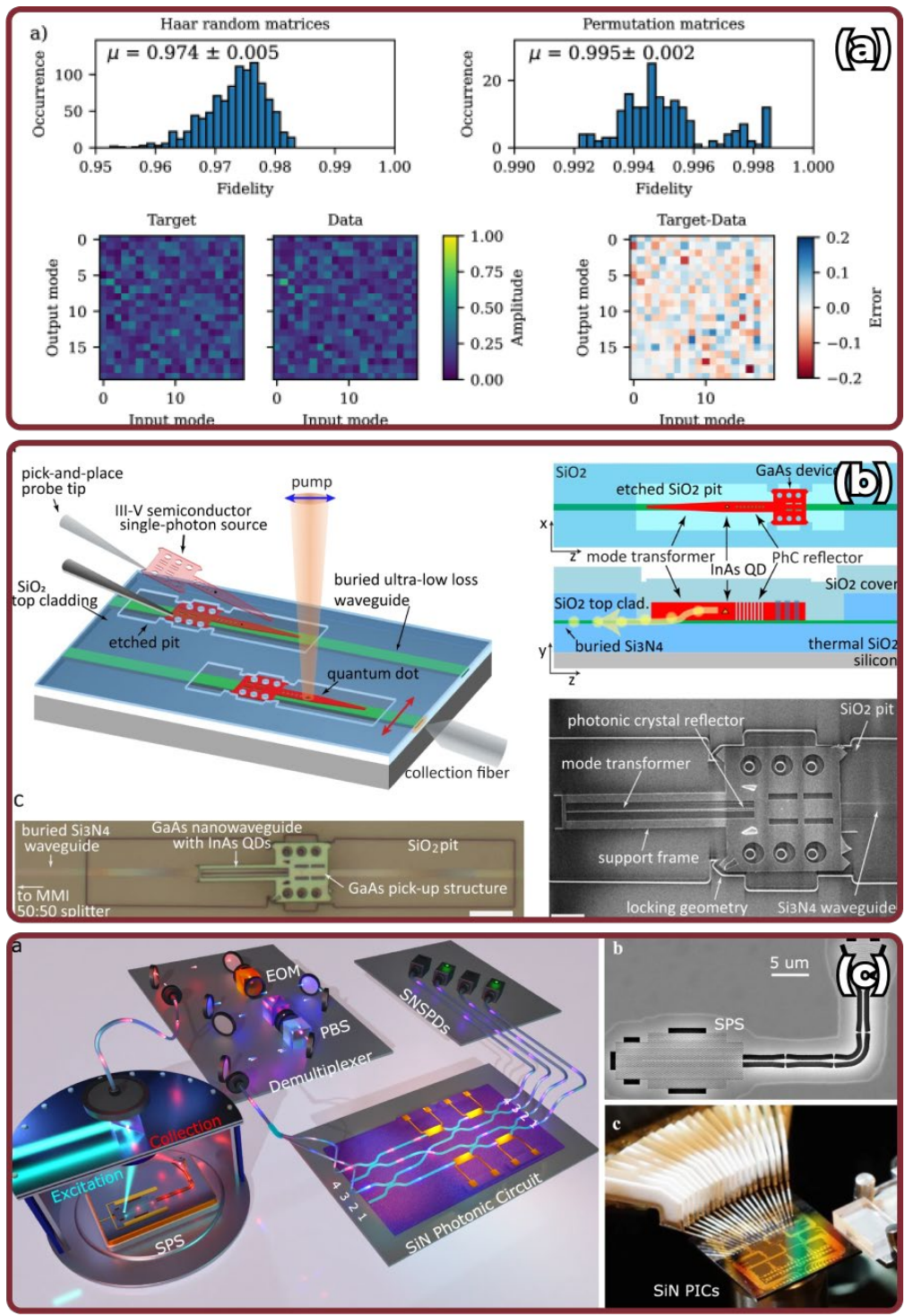

**Figure 19**. Silicon nitride photonics for quantum applications: (a) classical and quantum characterization of 20-mode reconfigurable quantum photonic processor in the C-band, reprinted from [7], under a Creative Commons license

(https://creativecommons.org/licenses/by/4.0/); (b) $Si_3N_4$ quantum photonic circuit with integrated InAs single-photon source, reprinted from [53], under a Creative Commons license (https://creativecommons.org/licenses/by/4.0/); (c) Interconnection of quantum dot single-photon source, low-loss $Si_3N_4$ photonic circuits, and SNSPD into a quantum photonic platform, reprinted from [263], under a Creative Commons license (https://creativecommons.org/licenses/by/4.0/).

Silicon nitride linear optical circuits could be a promising platform for neuromorphic computing due to low losses, as realized network accuracy reduces with losses growth during light propagation in PIC elements [265].

There are three most popular photonic platforms based on PICs: first one is based on a mesh of interconnected waveguides that is used for matrix multiplication [266,267], second one is based on parallelized computing using chip-based optical frequency combs [165] and the last one on incoherent crossbar arrays [149]. The first approach promises better scalability as information is coded in light channels, and their number increase is limited only by circuit size. Mesh of interconnected waveguides that is used for matrix multiplication of input signals, which is the most power-consuming operation for all types of neural networks. This type of architecture utilizes phase control through the use of integrated MZI structures cascaded together with electronically controllable phase shifters. Most promising solutions were previously demonstrated for silicon waveguides. Proof-of-principle demonstration allowed vowel recognition using 4x4 waveguides mesh [267]. For silicon nitride platform four input/output devices, shown in **Figure 20**a, were fabricated by LioniX International in a multi-project wafer run based on the silicon nitride TriPleX asymmetric double-strip waveguide platform [266]. The fabricated device study demonstrated its suitability to analogically compute photonic matrix–vector multiplications. For further photonic neuromorphic networks size growth silicon nitride looks as a promising platform, which would increase operations accuracy due to its low losses. However, on-chip training has not yet been shown for this platform, as it requires fast optical modulators implementation. As was previously mentioned in Section 3.1.3, most of the state-of-the-art silicon nitride circuits use heaters for optical modulation. Thus, circuits suffer from limited accuracy, high power consumption and low speed of operations. Recently, several works show application of opto-electro-mechanical (OEM) phase shifters for low insertion loss modulators in neuromorphic PICs [268–270]. OEM phase shifters have low power consumption compared to thermo-optic phase shifters, low insertion losses compared to electro-optic phase shifters and sufficient tuning speed in MHz range. On $Si_3N_4$ photonics platform, such phase shifters ensure losses of 0.5 dB with a high extinction ratio of 31dB and a maximum phase shift of 13.3 $\pi$ [268]. Similar type of device has been demonstrated for PIC on SOI platform [270]. It was proposed for performing linear optical transformations for general matrix-matrix multiplication with array sizes up to 128x128 [271]. Also, the advantage of OEM phase shifters is the possibility cryogenic temperatures operating, which opens up the possibility of use in quantum neuromorphic computing. Another optimization possibility is multimode interferometers (MMI) usage, which could be an alternative for optical convolution processing [272,273]. In recently proposed work [273], processing unit consist of two MMIs with programmable phase shifter array, into which incoherent lights with different wavelengths are fed. Such configuration allows for parallel processing of three convolutional computing operations [273].

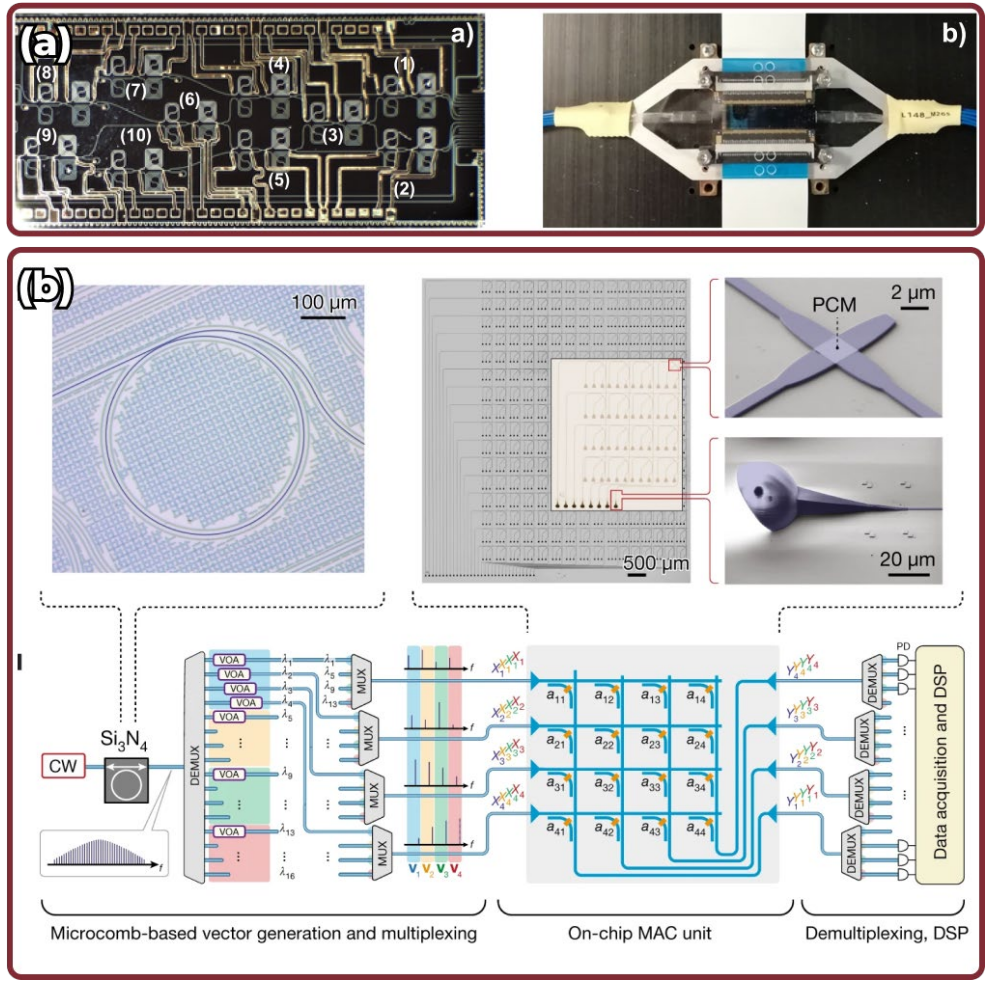

**Figure 20**. Neuromorphic processing on reconfigurable integrated optics: (a) Thermally tuned SiN processor for convolutional neural network, reprinted from [266], under a Creative Commons license (https://creativecommons.org/licenses/by/4.0/); (b) All-optical integrated photonic hardware accelerator for image processing, reprinted with permission from [the Licensor]: [Nature][165], [COPYRIGHT] (2021).

The second approach for photonic neuromorphic circuit fabrication is based on parallelized computing using chip-based optical frequency combs [Figure

20(b)] [165,274]. It is used to code signal in different wavelengths, which together with channels growth could raise network dimension to desirable values for practical applications. Silicon nitride is a promising platform which was previously used for optical frequency combs implementation [275]. The recently demonstrated photonic tensor core based on this technology operated at a speed of two trillion ($10^{12}$) MAC operations per second. This work could be positioned as proof-of-principle, but it is still far away from real neuromorphic device [265]. The last approach uses incoherent crossbar array [149,227] for matrix-vector multiplications with wavelength division multiplexing (WDM). [154,246]. For this architecture, information is encoded in the intensity of the optical pulses at different wavelengths, and coupling of each input of array structure to every output is precisely and separately controlled with on-chip high-speed variable attenuators [276]. This concept is especially applicable for constant matrices processing (i.e., edge detection [165]) and could be further improved with implementation of hybrid electro-optic hardware accelerators [265,277].

## 4. Challenges and outlook

In this review, we present the current state of silicon nitride integrated photonics. The level of passive components fabrication technology, as well as active devices integration, is discussed, considering recently presented advanced photonic-based devices in a variety of applications.

With the significant progress described above, high-performance silicon nitride photonics conquers the market in life sciences, automotive navigation and critical infrastructure security, metrology and sensing, quantum and neuromorphic computing. Most discrete devices previously implemented using bulk optics and optical fibers can be easily fabricated on a chip because of the ultralow losses in the $Si_3N_4$ waveguides and the high level of laser and photodetectors integration techniques. Silicon nitride PICs production is already widely spread by research groups and several photonic foundries, including LioniX, IMEC, Tower, AIM and Ligentec [278], reaching wafer diameters up to 8 inches [279]. Continuous improvement in PIC technology opens new frontiers and promises scalable, low-cost, and high-volume production.

However, many challenges remain before $Si_3N_4$-based photonic devices can be widely available. Along with the high level of technology, the problems of waveguides surface roughness and light absorption remain relevant. The need for long and high-temperature annealing restricts its compatibility with CMOS technology [69]. Continued work on the application of new precursors and deposition techniques [121,123,124] could provide more flexibility to technology with an affordable loss level. Although much progress has been made against surface roughness [46,50,99], photonic circuits still suffer from high sidewall scattering. To bring losses to the truly absorption-limited level, further research needs to be conducted on new techniques of sidewall roughness reduction to obtain sub-0.1 nm RMS values. Other challenges to be solved include the low index contrast in silicon nitride waveguides compared to SOI, the coupling efficiency of grating and edge couplers [132,143]. Achievement of low coupling losses nowadays requires complex and time-consuming component design [140,280], as well as additional process steps, which are currently underdeveloped to ensure mass production [60,134,135,156].

A further challenge is posed by the continuing growth of data transfer and processing. Recent advances in silicon nitride photonics have focused on passive applications and the main barrier here is the ability to control optical signals using only thermo-optics [202]. Although significant work has been done on material choice for on-chip electro-optics [169,210], an efficient demonstration of high-rate modulation on a silicon nitride platform is still not available. A possible way to overcome such challenges is to focus on the research and development of new materials [213] with further progress in integration methods with mature SOI and thin-film LNOI electro-optic solutions [207,211]. Recently, opto-electro-mechanical (OEM) phase shifters for low insertion loss modulators on $Si_3N_4$ photonics platform were shown. They represent a compromise solution with MHz frequency range, low insertion losses of 0.5 dB and a high extinction ratio of 31dB [219]

It is difficult to overestimate the contribution of silicon nitride photonics in telecommunications and nonlinear optics, through which the unique characteristics of integrated light sources have been achieved [93,234]. Meanwhile, the development of integrated photonics for the visible, near- and mid-IR wavelength bands is near the initial level. The need to reduce the structures size dramatically increases the complexity of the technology, and consequently, the losses in the devices [85,99]. The absence of high-performance integrated lasers and photodetectors prevents the manufacturing of complex and low-footprint devices [25,172]. Although there is still much to be done in technology and photonics infrastructure, the foundation laid in telecom will ensure subsequent progress for new application-specific markets, including quantum technologies and neuromorphic computing due to low losses, as well as personalized medicine devices and LiDAR sensors due to wide wavelength bandwidth [51,174,176].

In conclusion, we hope that our work will be a useful resource for the integrated photonics community. We ensure that silicon nitride photonics will greatly benefit from further close cooperation between R&D laboratories and photonic foundries. Continued research efforts in the field of new materials, technology, and multilayer integration will allow for large-scale, low-cost, and energy-efficient systems based on a heterogeneously integrated photonics platform that is ready for market and novel applications that are currently even difficult to predict.

**Conflict of Interest**

The authors declare no conflicts of interest.

## Keywords

silicon nitride photonics, light detection and ranging, quantum photonic computing, neuromorphic computing, optical frequency combs

## Biographies

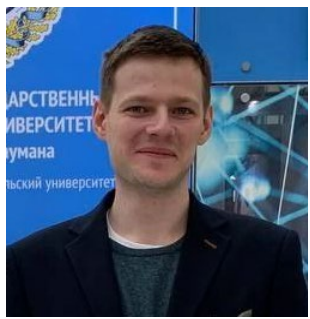

**Kirill Alekseevich Buzaverov** received the B.S. degree in material science in 2019 and the M.S. degree in high-tech plasma and power plants in 2021 from the Bauman Moscow State Technical University (BMSTU, Russia). He joined FMN Laboratory at BMSTU and Dukhov Research Institute of Automatics as a researcher in 2019, where he is currently working towards the Candidate of Science degree. His research interests are primarily concerned with integrated photonics and its applications in the fields of quantum computing and light detection and ranging.

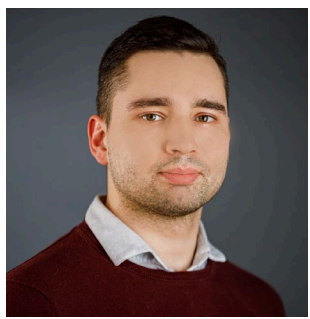

**Aleksandr Sergeevich Baburin** is Team Leader in the FMN Laboratory at Bauman Moscow State Technical University (BMSTU, Russia) and Dukhov Research Institute of Automatics, where he works on photonic integrated circuits for high performance computing, sensorics and LIDAR applications. He received his M.S. degree in micro and nanotechnology from the BMSTU in 2015. Subsequently in 2019 he received his Engineering Doctorate (Candidate of Science) in semiconductor technology from the BMSTU. From 2014 to 2018 he worked as a research engineer in the field of thin films deposition in the Research and Educational Center Functional Micro/Nanosystems (FMN

Laboratory). His research interests include the fabrication of photonics devices, integrated photonics and thin films.

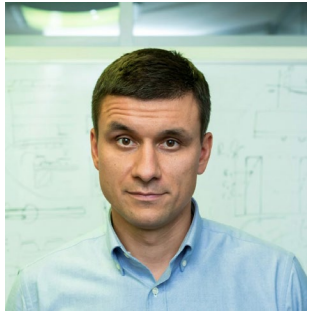

**Ilya Anatolievich Rodionov** is assistant professor at the Bauman Moscow State Technical University (BMSTU, Russia), received his PhD in semiconductor technology at BMSTU in 2010. Between 2005 and 2013 he was team leader on CMOS IC projects concerned with the development of sub-wavelength lithography, photomask design and OPC implementation at the Scientific Research Institute of System Analysis of Russian Academy of Science (SRISA RAS), Moscow. Currently, he is founding director of the Research and Educational Center Functional Micro/Nanosystems (FMN Laboratory) at BMSTU and Dukhov Research Institute of Automatics. He is working on design and nanofabrication technology of nanophotonics and optics, quantum simulators and sensors, lab-on-chip, photonics and MEMS sensors, and plasmonic devices.